\documentclass[aps,prx,reprint,superscriptaddress,balancelastpage,longbibliography]{revtex4-2}
\usepackage{graphicx}
\usepackage{multirow}
\usepackage{amsmath,amssymb,amsfonts}
\usepackage{amsthm}
\usepackage{mathrsfs}
\usepackage[title]{appendix}
\usepackage{xcolor}
\usepackage{textcomp}
\usepackage{algpseudocode}
\usepackage{siunitx}
\usepackage{listings}
\usepackage[english]{babel}
\usepackage[normalem]{ulem}
\makeatletter
\AtBeginDocument{
  \g@addto@macro{\appendix}{\renewcommand{\p@subsection}{\@Alph\c@section}}
}
\makeatother
\begin{document}

\title{Exploring dynamics of individual vortices in a superconductor via a levitated magnetic transducer}

%%%%%%%%%%%%%%%%%%%%%%%%%%%%%%%%%%%%%%%%%%%%%%%%%%%%%%%%%%%%%%%%%%Author%%%%%%%%%%%%%%%%%%%%%%%%%%%%%%%%%%%%%%%%%%%%%%%%%%%%%%%%%%%%%%%%%%%%%%%
\author{Yiqi~Wang}
\thanks{These authors contributed equally to this work}
\affiliation{Department~of~Physics,~Harvard~University,~Cambridge,~MA~02138,~USA}

\author{Trisha~Madhavan}
\thanks{These authors contributed equally to this work}
\affiliation{Harvard~John~A.~Paulson~School~of~Engineering~and~Applied~Sciences,~Harvard~University,~Cambridge,~MA~02138,~USA}

\author{J.~DaLi~Schaefer}
\thanks{Current address: Rigetti Computing, Berkeley, California, 94710, USA.}
\affiliation{Department~of~Physics,~Harvard~University,~Cambridge,~MA~02138,~USA}

\author{Addison~NewRingeisen}
\affiliation{Quantum~Science~and~Engineering,~Harvard~University,~Cambridge,~MA~02138,~USA}

\author{Frankie~Fung}
\thanks{Current address:Applied Materials, Inc., Santa Clara, California, 95051, USA.}
\affiliation{Department~of~Physics,~Harvard~University,~Cambridge,~MA~02138,~USA}

\author{Mikhail~D.~Lukin}
\email{lukin@physics.harvard.edu}
\affiliation{Department~of~Physics,~Harvard~University,~Cambridge,~MA~02138,~USA}

%%%%%%%%%%%%%%%%%%%%%%%%%%%%%%%%%%%%%%%%%%%%%%%%%%%%%%%%%%%%%%%%%%Abstract%%%%%%%%%%%%%%%%%%%%%%%%%%%%%%%%%%%%%%%%%%%%%%%%%%%%%%%%%%%%%%%%%%%%%%%
\begin{abstract}
    Trapped vortices determine fundamental properties of superconductors and play an important role in many practical applications such as magnetic levitation, however their complex dynamics remain poorly understood. Here, we use the mechanical motion of micron-scale levitated magnetic particles to probe the dynamics of individual vortices. Specifically, we show that the dynamics of levitated magnets are strongly influenced by vortices trapped in the YBCO superconducting film.
    We observe random telegraph signals in the mechanical frequency, dissipation rate, and energy of levitated particles, which we attribute to random tunneling of individual vortices. The nonlinearity of vortex-defect interaction manifests as non-exponential decay in ringdown measurements, revealing a complex underlying potential landscape. 
    Our results provide insights into elusive dissipation mechanisms in superconducting levitated systems, open new avenues for using levitated magnets as sensitive probes of static and dynamic properties of individual vortices in superconductors and their interactions with material disorder, and point toward novel routes for using magnetic particles as highly coherent mechanical transducers.
\end{abstract}
%%%%%%%%%%%%%%%%%%%%%%%%%%%%%%%%%%%%%%%%%%%%%%%%%%%%%%%%%%%%%%%%%%Introduction%%%%%%%%%%%%%%%%%%%%%%%%%%%%%%%%%%%%%%%%%%%%%%%%%%%%%%%%%%%%%%%%%%

\maketitle
\section{Introduction}
Coherent manipulation of the mechanical motion of levitated macroscopic objects is an exciting frontier in science and engineering. Precise control and measurement of levitated system dynamics facilitates the search for new physics~\cite{gonzalez_2021_levitodynamics_science,blakemore_2021_search_PRD,moore_2014_search_PRL,wang_2024_mechanical_PRL,afek_2022_coherent_PRL}, opens new avenues for quantum transduction~\cite{gieseler_2020_single_PRL} and sensing applications~\cite{ranjit_2016_zeptonewton_PRA,ahn_2020_ultrasensitive_NatNano,timberlake_2019_acceleration_APL}. Although optical levitation has already enabled ground-state cooling of sub-micrometer particles~\cite{delic_2020_cooling_nature, magrini_2021_real_science} and the creation of delocalized quantum states~\cite{Rossi_2025_quantum_prl,kamba_2025_quantum_science}, there is great interest in extending these results to larger particles and longer coherence times~\cite{bose_2017_spin_PRL}. 

Magnetic levitation associated with superconducting systems is a fundamental phenomenon with many practical applications~\cite{werfel_2012_superconductor_SST,bernstein_2020_superconducting_SST,ahrens_2025_levitated_PRL}. While magnetic levitation is a promising approach to realize stable levitation of millimeter-sized magnetic particles both at room temperature~\cite{lewandowski_2021_high_PRApplied, Tian_2024_feedback_APL} and cryogenic temperatures~\cite{gieseler_2020_single_PRL, hofer_2023_high_PRL}, dissipation mechanisms of magnetically levitated particles are not well understood, with mechanical quality factors limited to $10^7$~\cite{hofer_2023_high_PRL}. 
A deeper understanding of the dissipation mechanisms \cite{gieseler_2020_single_PRL, gutierrez_2023_superconducting_PRApplied,smit_2025_superconducting_arXiv} can provide novel insights into the properties of superconducting systems and is critical for extending coherent control of magnetically levitated particles into the quantum regime. Specifically, in the case of a permanent magnet levitated over a type-II superconductor, magnetic fields enter superconductors as quantized vortices, each carrying one flux quantum $\Phi_0=h/2e$. Trapped vortices change the magnetic response of superconductors, but this response lacks systematic studies~\cite{gieseler_2020_single_PRL, kordyuk_1998_magnetic_JAP}. Here, we experimentally investigate the dynamics of a micron-scale magnet levitated above the superconductor over a wide range of temperatures, oscillation amplitudes, and magnetic field configurations, intending to characterize the influence of trapped magnetic vortices on the magnet's dynamical behavior. We show that vortex dynamics affect the resonance frequencies, dissipation, and energy of a levitated magnetic oscillator.  We observe random telegraph signals in the mechanical frequency, dissipation rate, and energy of levitated particles which we attribute to random tunneling of individual vortices. These results provide new insights into a long-standing question regarding dissipation mechanisms in superconducting levitated systems and point toward new avenues for harnessing magnetic particles as highly coherent mechanical transducers~\cite{gieseler_2020_single_PRL,fung_2024_toward_PRL}. 

    %%%%%%%%%%%%%%%%%%%%%%%%%%%%%%%%%%%%%%%%%%%%%%%%%%%%%%%%%%%%%%%%%%%%%%%%Figure_1%%%%%%%%%%%%%%%%%%%%%%%%%%%%%%%%%%%%%%%%%%%%%%%%%%%%%%%%%%%%%%%%%%%%%%%%%%%%
    \begin{figure}
        \centering  
        \includegraphics[width=1\linewidth]{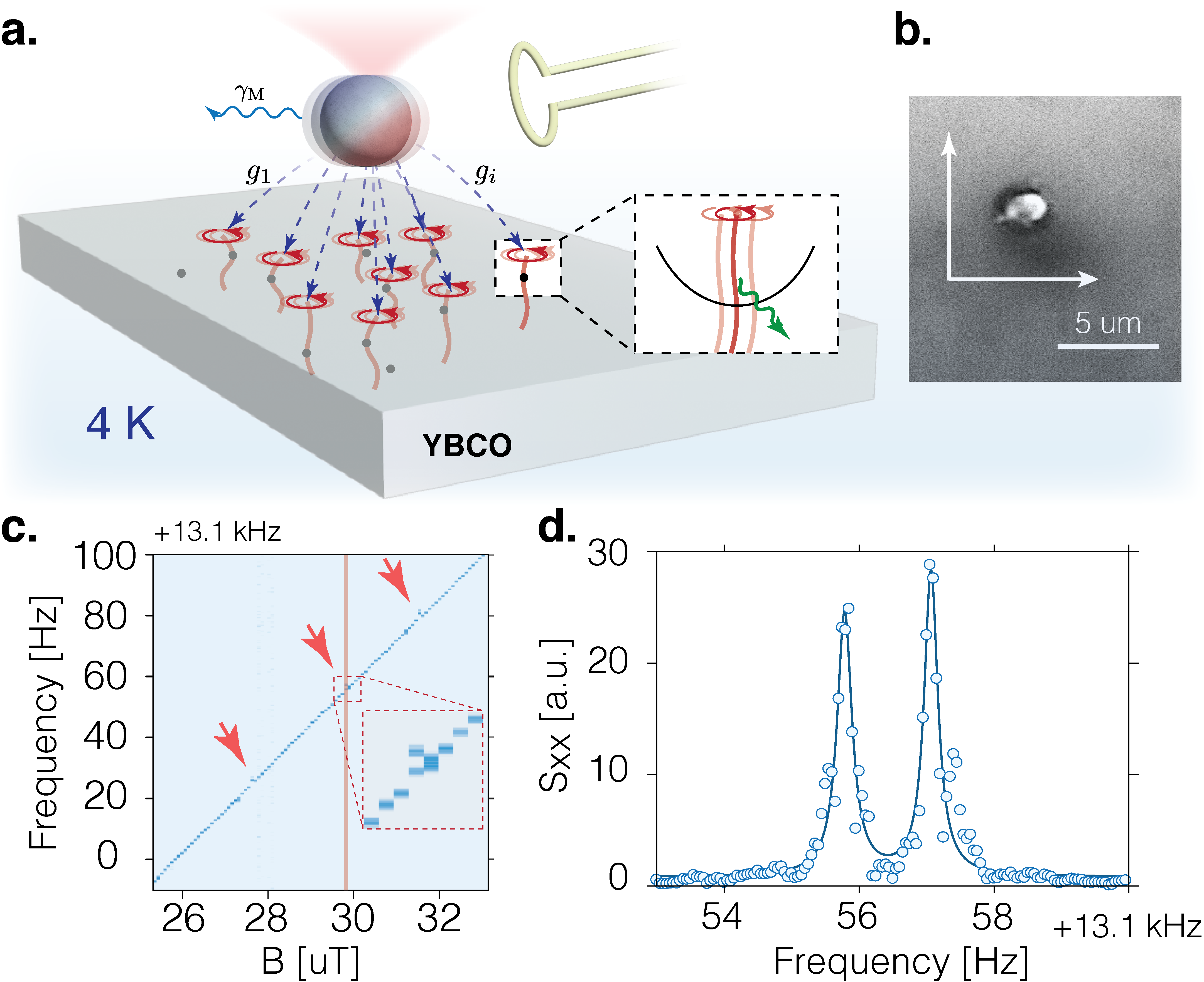}
        \caption{\textbf{Experiment setup and mechanical properties.}~
        (a) A micrometer-scale spherical magnet is trapped above the YBCO sample in a cryostat at \SI{6}{\kelvin}. The motion of the magnet is measured by a tightly focused \SI{637}{\nano\meter} laser and is excited by magnetic fields from a nearby coil. The magnet couples to individual vortices that are pinned by material defects. 
        (b) A microscope image of the levitated magnet. The coordinate axis represents the lab frame used in angle measurements.
        (c) Spectrum of $y$ mode during a fine magnetic field sweep. Red arrows highlight discontinuities in resonance frequency response.  
        (d) 
        An example of a double-peak mechanical spectrum at $B$ highlighted by the red shaded area in (c).   
    }
        \label{fig:1}
    \end{figure}
    %%%%%%%%%%%%%%%%%%%%%%%%%%%%%%%%%%%%%%%%%%%%%%%%%%%%%%%%%%%%%%%%%%%%%%%%%%%end%%%%%%%%%%%%%%%%%%%%%%%%%%%%%%%%%%%%%%%%%%%%%%%%%%%%%%%%%%%%%%%%%%%%%%%%%%%%
\section{Experimental setup and observations}

\subsection{Levitated magnetic oscillators}
    Our experiments utilize a \SI{1.5(1)}{\micro\meter} radius Nd-Pr-Fe-Co spherical magnet levitated above a \SI{270}{\nano\meter} thick single-crystal $\mathrm{Y Ba_{2} Cu_{3} O_{7}}$~(YBCO) film, as shown in Fig.~\ref{fig:1}(a). The YBCO sample is grown on a \SI{500}{\micro\meter} thick lanthanum aluminate substrate and placed in a cryostat with the lattice $c$-axis along $\hat{z}$. The YBCO is field cooled such that magnetic fields from the magnet partially penetrate the superconductor, enabling levitation. All six degrees of freedom, three translational and three rotational, are effectively harmonic oscillators~\cite{gieseler_2020_single_PRL, kordyuk_1998_magnetic_JAP}.

    We optically probe the motion of the magnet using a \SI{637}{\nano\meter} focused laser~\cite{hansen_2025_optical_arXiv}. To minimize heating effects, the incident laser power is kept below \SI{300}{\pico\watt}, corresponding to a displacement sensitivity up to \SI{0.4}{\nano\meter\per\sqrt{Hz}}. All mechanical modes are identified by sharp peaks in the photon count spectrum. 
    The levitation and measurement protocols are detailed in Appendix~\ref{sec:Levitation protocol}. In this work, we focus primarily on two in-plane translation modes,  $x$ and $y$, with frequencies between \SI{5}{\kilo\hertz} and \SI{30}{\kilo\hertz}. Additionally, a pair of magnetic coils outside the cryostat allows us to apply a homogenous magnetic field $B$ to the levitated magnet, rotating its magnetization axis such that all resonance frequencies depend on $B$, as shown in Fig.~\ref{fig:1}(c). Surprisingly, several discontinuities in mechanical resonance frequency are observed during the $B$ sweep, indicated by red arrows. One example is shown in Fig.~\ref{fig:1}(d), revealing a double-peak that cannot be solely described by net effects of surface currents.
    %%%%%%%%%%%%%%%%%%%%%%%%%%%%%%%%%%%%%%%%%%%%%%%%%%%%%%%%%%%%%%%%%%%%%%%%%%%Figure_2%%%%%%%%%%%%%%%%%%%%%%%%%%%%%%%%%%%%%%%%%%%%%%%%%%%%%%%%%%%%%%%%%%%%%%%%%%%%
    
    \begin{figure*}
        \centering
        \includegraphics[width=0.7\linewidth]{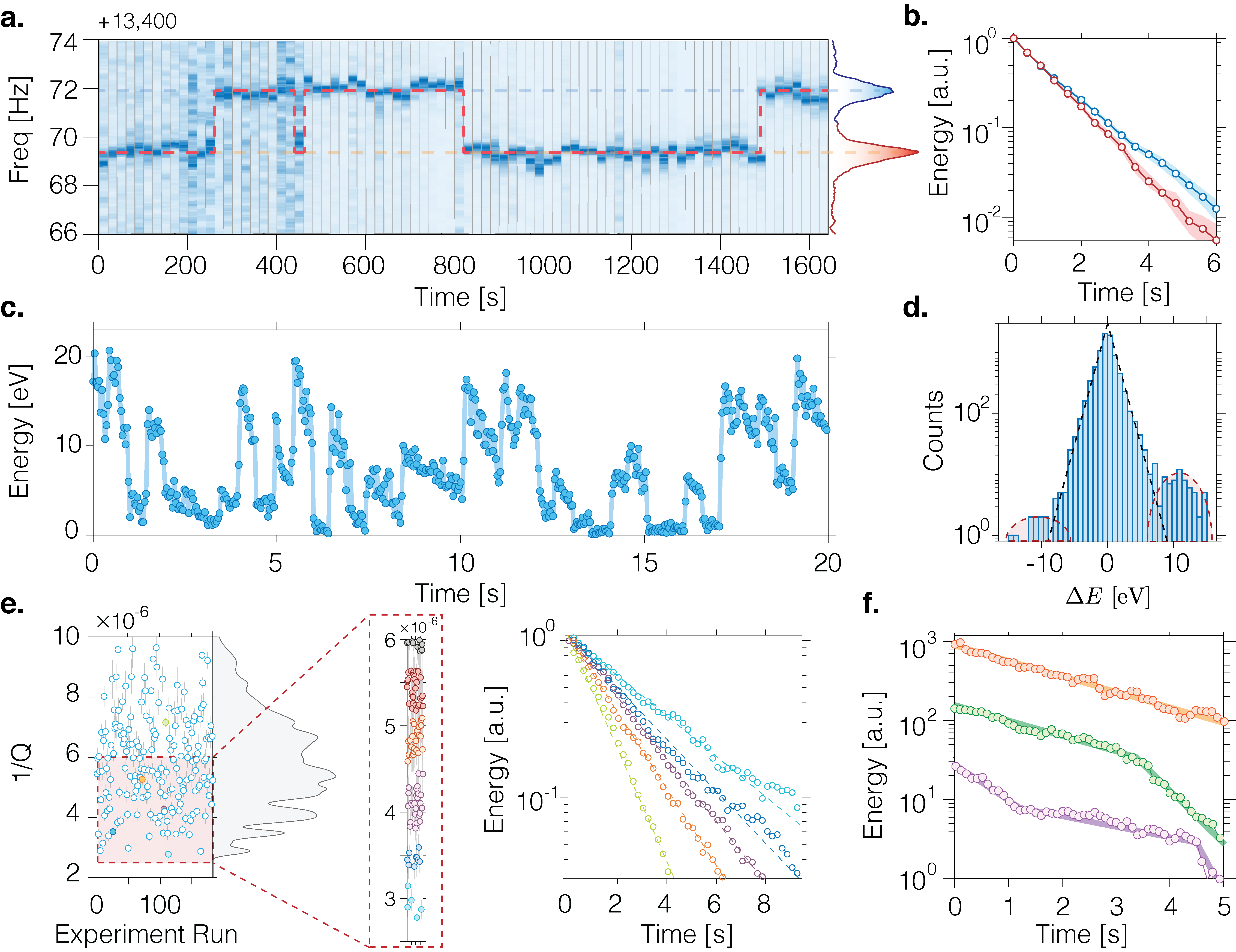}
        \caption{\textbf{Random telegraph signals in mechanical dynamics.}~
        (a) A continuous measurement of the mechanical spectrum at $B=$~\SI{53}{\micro\tesla}. The red dashed line highlights the resonance frequency. The average spectrum is shown on the right edge, featuring two distinct peaks. 
        (b) Averaged repeated ringdown measurements categorized based on the frequency at the beginning of the ringdown, where colors of circles correspond to peaks with the same color in (a). Shaded areas are 1 s.d. determined from statistics of repeated data. 
        (c) A real-time trace of kinetic energy at $B=$~\SI{34}{\micro\tesla} with each circle taken from a \SI{40}{\milli\second} measurement. 
        (d) Histogram of energy differences between each two consecutive points in (c). Black dashed lines mark the exponential distribution of thermal motion. Two pink bumps highlight anomalous events.
        (e) Inverse mechanical quality factors $1/Q$ from repeated ringdown measurements with identical initial conditions. Each data point is obtained by fitting a ringdown to a simple exponential decay, with error bars representing 2 s.d. from the fit. The gray shaded area is the distribution using a Gaussian kernel density estimator (see Appendix~\ref{sec:Gaussian_kernel_density_estimator}).
        The zoomed-in plot highlights the clustering of dissipation rates for the high-Q region. Representative ringdowns are shown on the right, where the color of each ringdown refers to the histogram data point of the same color.  
        (f) Examples of three ringdown measurements at $B=$~\SI{-362}{\micro\tesla}, during which damping rates suddenly jump.}
        \label{fig:2}
    \end{figure*}
    %%%%%%%%%%%%%%%%%%%%%%%%%%%%%%%%%%%%%%%%%%%%%%%%%%%%%%%%%%%%%%%%%%%%%%%%%%%%end%%%%%%%%%%%%%%%%%%%%%%%%%%%%%%%%%%%%%%%%%%%%%%%%%%%%%%%%%%%%%%%%%%%%%%%%%%%%

\subsection{Telegraph signals in particle dynamics}
    To further investigate abnormal observations, we systematically study dynamics under different conditions.
    At a fixed magnetic field $B=\SI{53}{\micro\tesla}$, a time trace of the mechanical spectrum (Fig.~\ref{fig:2}(a)) shows the mechanical resonance frequency randomly jumps between two distinct values. An averaged spectrum of the entire measurement on the side features a double-peak spectrum similar to Fig.~\ref{fig:1}(d) with \SI{2.8(1)}{\hertz} frequency separation. Similar frequency jumps have also been observed in other Meissner-levitation experiments~\cite{gutierrez_2023_superconducting_PRApplied}. 
    %In most cases, we observe spectral full-width-half-maxima greater than the linewidth estimated from ringdown measurements.
    We categorize repeated ringdown measurements at the same field into two groups based on the initial resonance frequency and find that the two mechanical frequencies are associated with two distinct dissipation rates, as shown in Fig.~\ref{fig:2}(b). In another magnetic field configuration ($B=$~\SI{-362}{\micro\tesla}), we find the damping rate changes distinctly even during a single ringdown measurement as shown in Fig.~\ref{fig:2}(f), akin to observations in levitated superconductors~\cite{hofer_2023_high_PRL}.

    In a separate levitation attempt, we repeat over one thousand ringdown measurements at a fixed drive power from which nearly 200 traces can be fit with high confidence using a simple exponential decay.
    Measured inverse quality factors $1/Q$ are shown in Fig.~\ref{fig:2}(e), where each error bar represents 2 s.d. determined from the fits. The dissipation rates extracted from repeated measurements randomly vary over a fivefold range across nearly two days of continuous data collection. The dissipation rates strikingly cluster around certain discrete values. Representative ringdown examples are shown in Fig.~\ref{fig:2}(e).
        
    Finally, we investigate the energy of the system under a white noise drive at $B=$\SI{34}{\micro\tesla}. The kinetic energy is continuously monitored through spectral measurements every \SI{40}{\milli\second} ($\gg 1/\omega_{\mathrm{m}})$. Figure~\ref{fig:2}(c) shows an example time trace of the kinetic energy fluctuating around 7.6~eV. However, unlike the Brownian motion of a harmonic oscillator, occasional energy jumps are observed at a rate much faster than the coherence time (\SI{0.5(1)}{\second}). To further elucidate this behavior, a histogram of the energy difference between two consecutive \SI{40}{\milli\second} bins during a \SI{300}{\second} measurement is shown in Fig.~\ref{fig:2}(d). Two distinct peaks are seen symmetric about the expected exponential distribution of Brownian motion, with a mean value of $\pm$~10.9~eV. 

%%%%%%%%%%%%%%%%%%%%%%%%%%%%%%%%%%%%%%%%%%%%%%%%%%%%%%%%%%%%%%%%%%%%%%%Section 3%%%%%%%%%%%%%%%%%%%%%%%%%%%%%%%%%%%%%%%%%%%%%%%%%%%%%%%%%%%%%%%%%%%%%%%%%%%%%%%%%
\subsection{Relation to superconductor properties}
\begin{figure*}
        \centering
        \includegraphics[width=0.7\linewidth]{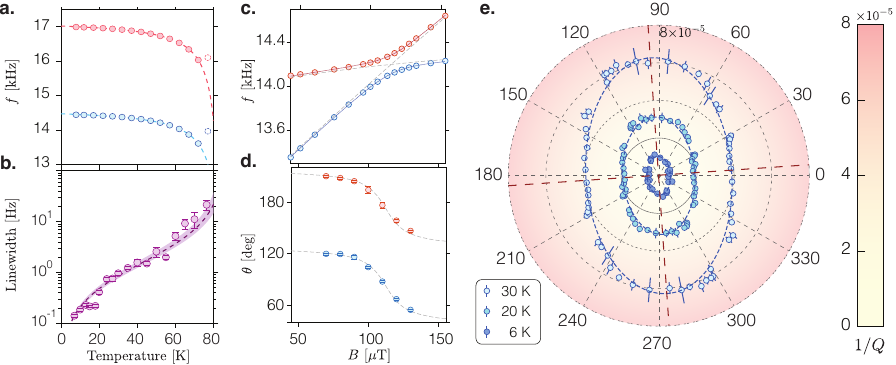}
        \caption{\textbf{Relations between mechanical dynamics and superconductor properties.}~
        (a) Mechanical resonance frequencies of the $x$ mode (red circles) and $y$ mode (blue circles) as a function of temperature. Corresponding dashed lines are fits to the model described in Appendix~\ref{sec:Mechanical-vortex interaction}. Dashed circles represent irreversible frequency changes at \SI{77}{\kelvin}.
        (b) Measured linewidth of the $y$ mode (purple circles) as a function of the YBCO temperature.
        Error bars are 2 s.d. determined from fits in ringdown measurements. The dashed line is a fit to the model detailed in Appendix~\ref{sec:Mechanical-vortex interaction} with the corresponding 1 s.d. (shaded purple area).
        (c) Mechanical frequencies of $x$ mode (red circles) and $y$ mode (blue circles) as a function of $B$. Grey dotted lines represent the expected avoided crossing from a linear coupled mode model. Grey dashed lines show the uncoupled mode frequencies. 
        (d) Corresponding measured angles of two in-plane normal modes in the lab frame during the $B$ field sweep. Grey dashed lines represent estimated angles due to mode hybridization. Error bars are 2 s.d. determined from fits in galvo scans to extract angles. 
        (e) Inverse quality factor $1/Q$ as a function of angles of two in-plane normal modes at 6~K, 20~K, and 30~K. Red dashed lines are YBCO in-plane crystal axes, extracted from XRD measurements (Appendix~ \ref{sec:YBCO characterization}).
        Error bars are 2 s.d. determined from fits. Dashed circles are elliptical global fits to extract the anisotropy of damping rates. The oscillation is bidirectional, so all data are duplicated and rotated by $180^\circ$ for clearer illustration.
        }
        \label{fig:3}
    \end{figure*}

    To investigate how superconducting properties influence the mechanical dynamics, we sweep the temperature of the YBCO from \SI{6}{\kelvin} to \SI{78}{\kelvin} and measure resonance frequencies and linewidths of mechanical modes. As shown in Fig.~\ref{fig:3}(a), increasing the superconductor temperature decreases the resonance frequencies of both modes.
    The amplitude of motion is kept below \SI{3.8}{\nano\meter} ($E<$~\SI{60}{\milli eV}) throughout the temperature sweep, and the frequency shifts are fully reversible over repeated sweeps. 
    When the temperature is raised close to $T_{\mathrm{c}}$, however, frequency changes become irreversible, as indicated by the dashed circles in Fig.~\ref{fig:3}(a). The mechanical dissipation rate also exhibits a pronounced temperature dependence, as plotted in Fig.~\ref{fig:3}(b).
    
    Furthermore, the angle of two in-plane normal modes can be precisely controlled. When two mode frequencies are tuned to be nearly degenerate by a homogeneous magnetic field, an effective linear coupling term leads to an avoided crossing in the mechanical mode spectrum shown in Fig.~\ref{fig:3}(c). The resulting mode hybridization changes the angle of each mode as expected in Fig.~\ref{fig:3}(d). A description of the linear coupled model is detailed in Appendix~\ref{sec: linear_coupling}. Interestingly, the mechanical dissipation rate exhibits a clear two-fold anisotropic angular dependence that persists at different temperatures, as shown in Fig.~\ref{fig:3}(e), where each point represents an average of 20 ringdown measurements. 
    The two-fold anisotropy, extracted by globally fitting three scaled ellipses to all measured results, is $1.62\pm0.08$~\cite{pereg_2004_absolute_PRB}. The fitted ellipses are rotated by $4.4^{\circ}\pm1.4^{\circ}$ with respect to the laboratory frame, coincident with the in-plane orientation of the YBCO lattice $(4^{\circ}\pm1^{\circ})$ obtained from XRD measurements detailed in Appendix~\ref{sec:YBCO characterization}. 
    
    %%%%%%%%%%%%%%%%%%%%%%%%%%%%%%%%%%%%%%%%%%%%%%%%%%%%%%%%%%%%%%%%%%%%%%%%%%%%Figure_3%%%%%%%%%%%%%%%%%%%%%%%%%%%%%%%%%%%%%%%%%%%%%%%%%%%%%%%%%%%%%%%%%%%%%%%%%%%%
    %%%%%%%%%%%%%%%%%%%%%%%%%%%%%%%%%%%%%%%%%%%%%%%%%%%%%%%%%%%%%%%%%%%%%%%%%%%%end%%%%%%%%%%%%%%%%%%%%%%%%%%%%%%%%%%%%%%%%%%%%%%%%%%%%%%%%%%%%%%%%%%%%%%%%%%%%

    Lastly, we repeat ringdown measurements with increasing initial energies. Figure~\ref{fig:4}(a) displays an energy-dependent non-exponential relaxation process at $B=$~\SI{92}{\micro\tesla}. An initial period of fast decay is followed by a sharp change to significantly slower dissipation. The time to reach the damping rate crossover $\tau_{\text{crs}}$ follows $\tau_{\text{crs}}=\tau_1\log(E_{\text{d}}/E^{\prime})$, where $\tau_1$ is the average decoherence time of the fast decay process, $E_{\text{d}}$ is the driven energy, and $E^{\prime}$ is a hypothetical threshold energy, indicating the existence of a threshold energy for a double-exponential decay form. Similar non-exponential decays have been observed in superconducting qubits related to quasiparticle population~\cite{gustavsson_2016_suppressing_Science}, and in nanomechanical resonators due to nonlinear mode coupling~\cite{guttinger_2017_energy_NatNan}, or interactions with a small number of two-level systems (TLS)~\cite{cleland_2024_studying_NatCom}. None of these mechanisms can reasonably explain all observations in this experiment. 
    We further find that the hypothetical threshold energy $E^{\prime}$ depends on $B$. As the field is reduced from \SI{100}{\micro\tesla} to \SI{75}{\micro\tesla}, the initially low-dissipation, single-exponential decay evolves into a double-exponential decay, before reverting to a single-exponential decay with an increased dissipation rate. Within the double-decay regime, $E^{\prime}$ increases monotonically with increasing $B$, as shown in Fig.~\ref{fig:4}(d). 
    For a different levitated magnet, the same initial dissipation process bifurcates into two different rates after the energy drops below a certain threshold, as shown in Fig.~\ref{fig:4}(e). This non-trivial behavior represents direct evidence of a bistable solution in magnet dynamics.

%%%%%%%%%%%%%%%%%%%%%%%%%%%%%%%%%%%%%%%%%%%%%%%%%%%%%%%%%%%%%%%%%%%Discussion%%%%%%%%%%%%%%%%%%%%%%%%%%%%%%%%%%%%%%%%%%%%%%%%%%%%%%%%%%%%%%%%%%%%%%%%%%
\section{Interpretation of complex dynamics}

    These observations indicate unusual dynamics driven by a complex potential beyond a simple harmonic trap. 
    First, the observed temperature-dependent resonance frequencies contradict the simplified frozen–image-dipole model, which assumes a temperature-independent boundary condition set by superconductor surface currents. Both the resonance frequency and the mechanical damping rate exhibit a pronounced temperature dependence around $T_{\mathrm{c}}$, suggesting the trap is directly influenced by superconducting properties. In addition, the observed anisotropic dissipation rate further rules out isotropic damping mechanisms such as gas damping, and cannot be explained by typical eddy current damping (Appendix~\ref{sec:eddy_current_estimate}). Finally, sharp jumps in magnet dynamics suggest coupling to a reservoir composed of discrete fluctuators.

    \subsection{Mechanical-vortex coupling}\label{sec:Mechanical-vortex coupling}
%%%%%%%%%%%%%%%%%%%%%%%%%%%%%%%%%%%%%%%%%%%%%%%%%%%%%%%%%%%%%%%%%%%%%%%%%%%%Figure_4%%%%%%%%%%%%%%%%%%%%%%%%%%%%%%%%%%%%%%%%%%%%%%%%%%%%%%%%%%%%%%%%%%%%%%%%%%%%
    \begin{figure*}
        \centering
        \includegraphics[width=0.85\linewidth]{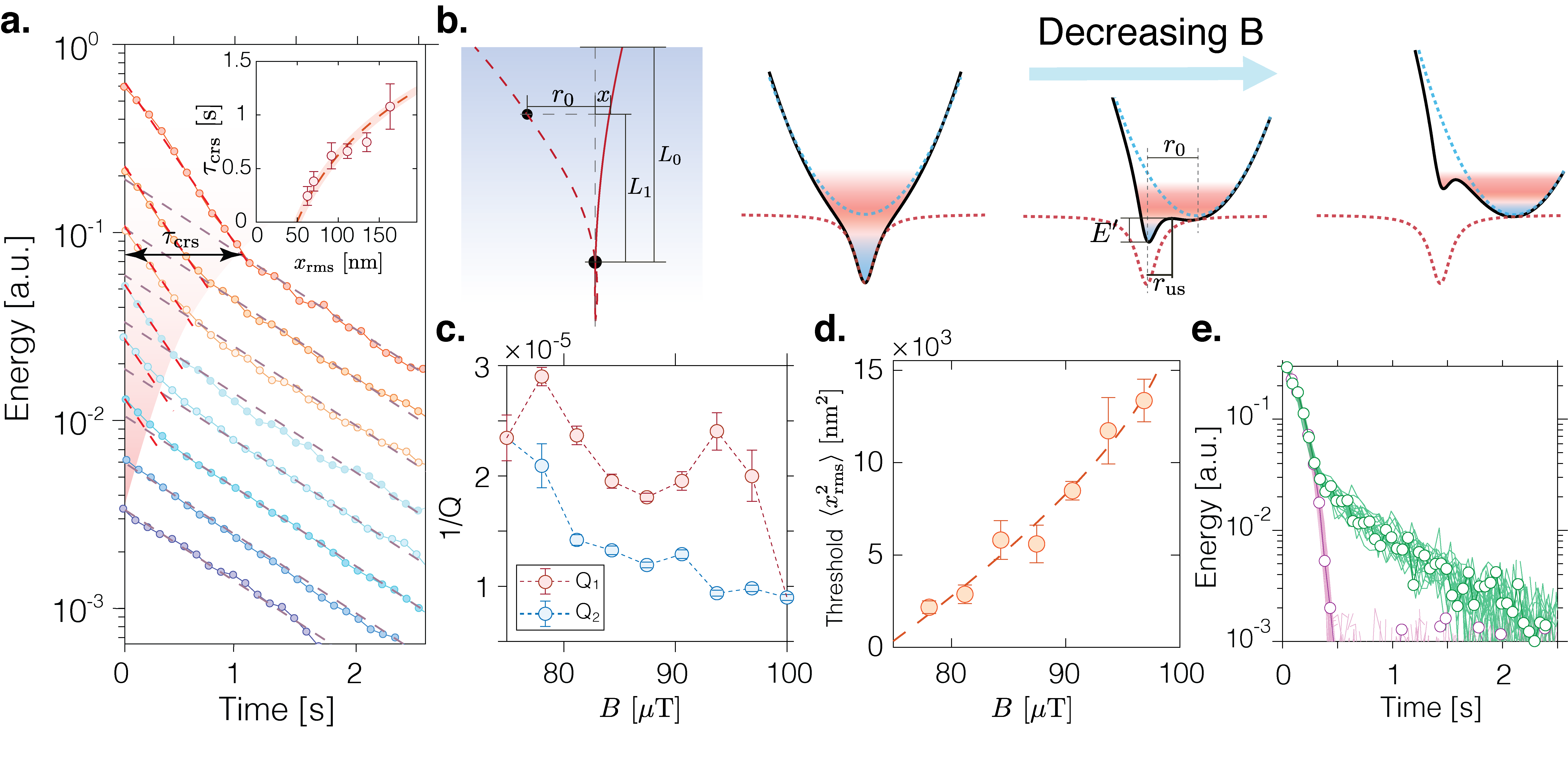}
        \caption{\textbf{Energy dependent relaxation.}
        (a) Averaged ringdown measurements at $B=$~\SI{92}{\micro\tesla} with an increasing starting amplitude from purple circles to red circles. Red and purple dashed lines depict fast decays in the beginning and the following slow damping, respectively. Inset: Damping rate crossover time $\tau_{\text{crs}}$ as a function of starting amplitudes. The dashed red line represents a fit to a hypothetical threshold energy. Error bars are 2 s.d. determined from uncertainties from fitting and statistical uncertainties.
        (b) (left) Schematic of a single vortex (red line) interacting with defects (black dots) in the superconductor. (right) Illustrations of hypothetical strong pinning potentials of a single vortex with decreasing $B$ field. The red dashed lines represent Lorentzian-like pinning potentials, the blue dashed lines are elastic potentials of a single vortex, and the black solid lines are the net Gibbs energy of a single vortex.
        (c) Inverse mechanical quality factors $1/Q$ as a function of $B$. Red circles and blue circles respectively denote quality factors of the initial fast damping and the following slow damping in double-exponential ringdowns. Error bars are 2 s.d. determined from uncertainties of the fittings. 
        (d) Corresponding threshold fluctuation amplitudes of double-exponential ringdowns. The orange dashed line is a fit detailed in Appendix~\ref{sec:Strong pinning theory}. Error bars are 2 s.d. determined from uncertainties of the fittings and statistical uncertainties.
        (e) An example of bifurcation in ringdown measurements. Solid purple and green lines are individual repeated ringdowns with distinct damping features. Circles highlight one example of each kind. 
        }
        \label{fig:4}
    \end{figure*}
    %%%%%%%%%%%%%%%%%%%%%%%%%%%%%%%%%%%%%%%%%%%%%%%%%%%%%%%%%%%%%%%%%%%end%%%%%%%%%%%%%%%%%%%%%%%%%%%%%%%%%%%%%%%%%%%%%%%%%%%%%%%%%%%%%%%%%%%

\begin{figure}
        \centering
        \includegraphics[width=0.9\linewidth]{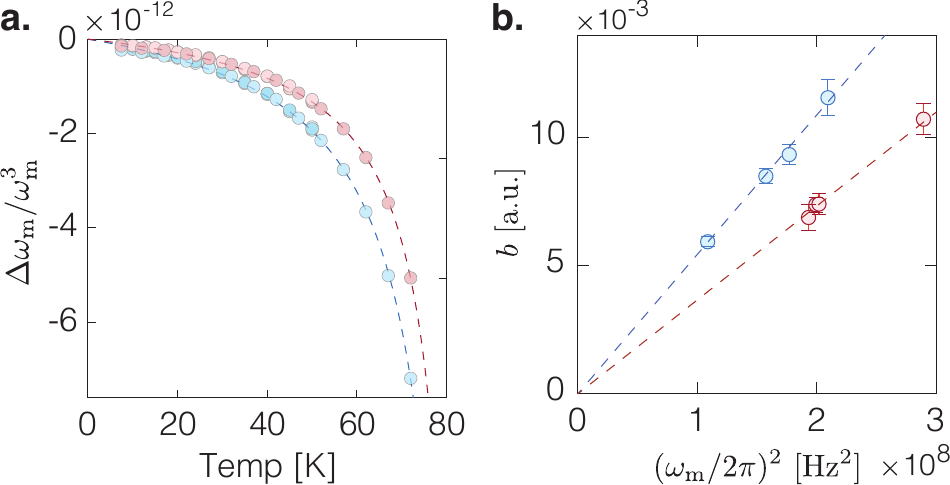}
        \caption{
        \textbf{Temperature dependence of mechanical resonance frequency. }
             (a) $\Delta \omega_{\mathrm{m}} / \omega_{\mathrm{m}}^3$ as a function of temperature for four different mode frequencies at \SI{6}{\kelvin}, represented by 
             red ($x$) and blue ($y$) circles with different shades. Dashed lines show the corresponding fits (see Appendix~\ref{sec:Temperature dependence at different B} for details)  
             (b) Fitted parameter $b$ as a function of $(\omega_{\mathrm{m}}/2\pi)^2$.
             Error bars represent 2 s.d. determined from the fits.   Dashed lines are linear fits.
        }
        \label{fig:5}
    \end{figure}

    We now argue that these observations can be explained by the interaction between the levitated magnet and individual trapped vortices. To quantitatively verify this hypothesis, we consider a simple mechanical-vortex model where each vortex is treated independently. These vortices, supported by surface supercurrents, behave like a collection of springs constraining the motion of the magnet. Vortices are typically modeled as one-dimensional classical elastic objects that interact with local pinning defects, such as impurities, vacancies, stacking faults, or grain boundaries~\cite{auslaender_2009_mechanics_natphy, buchacek_2019_strong_PRB}. 
    The interaction between each vortex and each pinning site $V_{\mathrm{p}}$ can be approximated as a harmonic potential, whose natural frequency is on the order of GHz~\cite{compton_2006_dynamics_PRL, nambisan_2025_quantum_arXiv, mehrnia_2024_observation_JMMM}. The effective spring constant of the pinning site can be characterized by the Labusch parameter, defined as $\alpha_{\mathrm{L}}=1/L \times\partial^2 V_{\mathrm{p}}/\partial s ^2$ with $L$ being the length of the vortex~\cite{labusch_1969_elastic_PSS, campbell_1969_response_JPC, doyle_1993_direct_PRL} and $s$ being the position of the vortex. Although levitated particles are physically isolated, their mechanical motion inevitably interacts with an ensemble of such far-off-resonant vortices. Phenomenologically, the dynamics of each degree of freedom can be described by the Hamiltonian 
    \begin{align} \label{eq:effective_coupled_hamiltonian}
        \mathcal{H}=\frac{1}{2}m\dot{x}^2+\frac{1}{2}m\omega^2_{\mathrm{m}}x^2+\sum_i V_{\mathrm{p,i}}(s_i)-\sum_{i}g_is_ix,
    \end{align}
    where $m$ is the mass of the magnet, $x$ is its displacement, $\omega_{\mathrm{m}}$ is the intrinsic mechanical resonance frequency, and $V_{\mathrm{p,i}}(s_i)$ describes the pinning potential of the \textit{i}-th vortex with a displacement $s_i$ from its equilibrium in a pinning center, which can be approximated as a harmonic potential of $k_{\mathrm{v,i}}s_i^2/2$.
     The interaction strength $g_i$ represents the effective force on the vortex by the magnet's displacement, which depends on the location of individual vortices and the magnetic moment of the levitated magnet. Although each vortex meanders through the superconductor and interacts with multiple pinning sites, we simplify the model by assuming that each vortex primarily couples to a single defect. As the magnet moves, the vortex will be displaced in phase with the far-off-resonant driving force from the magnet by $s_i\approx g_ix/k_{\mathrm{v,i}}$. This displacement results in an effective mechanical frequency $\omega_{\mathrm{m}}'$ given by
      \begin{align}\label{eq:freq_by_vortex}        
        \omega_{\mathrm{m}}'=\omega_{\mathrm{m}}(1-\frac{1}{2m\omega_{\mathrm{m}}^2}\sum\frac{g_i^2}{k_{\mathrm{v,i}}}).
      \end{align}
    Hence, the mechanical frequency decreases with the increasing superconductor temperature due to the softening of the vortex pinning potential, as shown in Fig.~\ref{fig:3}.
    The fits based on superconductor properties capture the measured temperature dependence, yielding an extracted $T_{\mathrm{c}}=$~\SI{85.8(9)}{\kelvin} that is in reasonable agreement with $T_{\mathrm{c}}=$~\SI{87(1)}{\kelvin} from resistance measurements~(see Appendix~\ref{sec:Temperature dependence}). 
    Noticing that $\sum g_i=m\omega_{\mathrm{m}}^2$, we have 
    \begin{align}\label{eq:freq_change_scale_freq}
        \frac{\Delta\omega_{\mathrm{m}}}{\omega_{\mathrm{m}}}\propto \frac{\omega_{\mathrm{m}}^2}{\bar{k}_{\mathrm{v,i}}(T)}
    \end{align}
    by treating each vortex identically,
    where $\Delta\omega=\omega_{\mathrm{m}}-\omega_{\mathrm{m}}'$ is the frequency change, $\bar{k}_{\mathrm{v,i}}(T)$ describes the mean temperature dependence of the vortex pinning. Figure~\ref{fig:5}(a) shows the normalized frequency change $\Delta\omega_{\mathrm{m}}/\omega_{\mathrm{m}}^3$ with increasing temperatures for different frequencies at 6~K. 
    Four sweeps exhibit nearly identical temperature dependence, consistent with the relation in Eq.~\eqref{eq:freq_change_scale_freq}. This relation is more evident in Fig.~\ref{fig:5}(b), where $b$ are extracted parameters from fits to $\Delta\omega_{\mathrm{m}}/\omega_{\mathrm{m}}=-b/\bar{k}_{\mathrm{v,i}}(T)$, showing a clear proportional relation to $\omega^2_{\mathrm{m}}$. A detailed description of the fit is given in Appendix~\ref{sec:Temperature dependence at different B}. We attribute the difference between $x$ and $y$ modes to the anisotropy of the pinning potential.
   
    Vortex motion involves the displacement of a nanometer-scale core containing normal electrons, thereby contributing to the dissipation of the system.
    In a phenomenological microscopic picture, this process can be viewed as an effective viscous drag force acting on a moving vortex, described by the Bardeen-Stephen viscosity~\cite{bardeen_1965_theory_PR}. 
    The pinning center softens with increasing temperature, resulting in increasing vortex displacement and higher mechanical dissipation rate. This hypothesis is further supported by the two-fold anisotropy of dissipation, closely aligned to the YBCO in-plane lattice. A likely origin is an anisotropic vortex-defect interaction, arising from the asymmetry of the $\operatorname{a\text{–-}b}$ axis in YBCO~\cite{nishizaki_2003_vortex_JLTP}, or directional defects such as twin boundaries and stacking faults~\cite{palau_2006_crossover_PRL}.
    The orientation of the damping anisotropy is also consistent with the result in Fig.~\ref{fig:5}, where a higher damping rate corresponds to a softer pinning potential.

     %%%%%%%%%%%%%%%%%%%%%%%%%%%%%%%%%%%%%%%%%%%%%%%%%%%%%%%%%%%%%%%%%%%end%%%%%%%%%%%%%%%%%%%%%%%%%%%%%%%%%%%%%%%%%%%%%%%%%%%%%%%%%%%%%%%%%%%
     %%%%%%%%%%%%%%%%%%%%%%%%%%%%%%%%%%%%%%%%%%%%%%%%%%%%%%%%%%%%%%%%%%%end%%%%%%%%%%%%%%%%%%%%%%%%%%%%%%%%%%%%%%%%%%%%%%%%%%%%%%%%%%%%%%%%%%%

    \subsection{Vortex-defects interaction}
    For the typical conditions used in our experiments, the number of vortices associated with the field from the magnet is on the order of tens (see Appendix~\ref{sec:Estimation_of_number_vortices}). We therefore attribute certain telegraph signals to the pinning-depinning of individual vortices.
    The dynamics of a vortex are governed by the competition between pinning forces and the elastic restoring force associated with vortex line tension, which together define an effective potential landscape. Considering a single vortex of length $L_0$ initially oriented along the ${z}$-axis, where bending the vortex increases the length, and hence its Gibbs free energy is given by 
    \begin{align}\label{eq:Gibbs_energy_pinning}
            G(x)=\frac{1}{2}\epsilon_{\text{L}}L_0\left(\frac{x}{L_1}\right)^2-\frac{V_{\text{p}}}{1+(x-r_0)^2/\epsilon^2}-\frac{\Phi_0B}{\mu_0}x,
    \end{align}
    where the first term represents the elastic potential that is proportional to the vortex length with $\epsilon_{\text{L}}$ setting the single-vortex energy scale~\cite{kuit_2008_vortex_PRB}. We model a pinning center by incorporating a Lorentzian potential
    at position $r_0$, with depth $V_{\text{p}}$ and width $\epsilon$~\cite{nambisan_2025_quantum_arXiv,buchacek_2019_strong_PRB}, sketched as red dotted curves in Fig.~\ref{fig:4}(b).
    The third term represents the Lorentz force exerted on the vortex by an in-plane magnetic field $B$, with $\Phi_0$ being the magnetic flux quantum. 
    Such an interaction can result in bistable solutions for vortex configurations related to unexpected behaviors in magnetic field sweeps as shown in Fig.~\ref{fig:1}~\cite{buchacek_2019_strong_PRB}. 
    Much like telegraph signals induced by TLSs in superconducting devices~\cite{zhang_2024_acceptor_PRL, faoro_2015_interacting_PRB, kennedy_2025_josephson_PRL} and nanomechanical resonators~\cite{maksymowych_2025_spectral_PRApplied}, we argue that observed jumps in the dynamics of the magnet are evidence of single vortex random pinning-depinning processes ~\cite{buchacek_2019_strong_PRB,nambisan_2025_quantum_arXiv}.  
    The reconfiguration of vortices, driven either by the external excitation or by thermally assisted creep, can modify the vortex stiffness and viscosity, leading to correlated shifts in the mechanical resonance frequencies and damping rates, as shown in Fig.~\ref{fig:2}(b).
    Vortex creeping further gives rise to the discrete dissipation rates observed in Fig.~\ref{fig:2}(e) and broadening of the mechanical spectrum~\cite{maksymowych_2025_spectral_PRApplied, kennedy_2025_josephson_PRL}.

    Accounting for such a non-harmonic potential of individual vortices in Eq.~\eqref{eq:Gibbs_energy_pinning} can further explain the nonlinear dissipation process.
    When the particle oscillates at high amplitudes, a vortex can detach from a pinning center, resulting in enhanced damping. As the oscillation amplitude decreases below a threshold, the vortex re-attaches to the defect, leading to slower energy decay. We observe two distinct decay rates in Fig.~\ref{fig:4}(a) that persist over a broad range of initial kinetic energies, suggesting that the dissipation in this case is dominated by a single vortex-defect interaction. 
    The magnetic field $B$ in Eq.~\eqref{eq:Gibbs_energy_pinning} effectively shifts $r_0$ by $\Delta r= \Phi_0BL_1^2/{\mu_0\epsilon_{\text{L}}L_0}$.
    As $r_0$ increases, the potential evolves from a deeply pinned regime to an anharmonic potential with metastable states, eventually leading to complete depinning from the pinning center, as illustrated in Fig.~\ref{fig:4}(c). 
   The measured threshold energy $E^{\prime}$ at different $B$ (Fig.~\ref{fig:4}(d)) quantitatively agrees with the prediction from Eq.~\eqref{eq:Gibbs_energy_pinning} (see Appendix~\ref{sec:Strong pinning theory}). Furthermore, a bistable double-well potential accounts for the nontrivial bifurcation observed in Fig.~\ref{fig:4}(e)~\cite{buchacek_2019_strong_PRB}.

    %%%%%%%%%%%%%%%%%%%%%%%%%%%%%%%%%%%%%%%%%%%%%%%%%%%%%%%%%%%%%%%%%%
    \begin{figure}[b]
        \centering
        \includegraphics[width=0.8\linewidth]{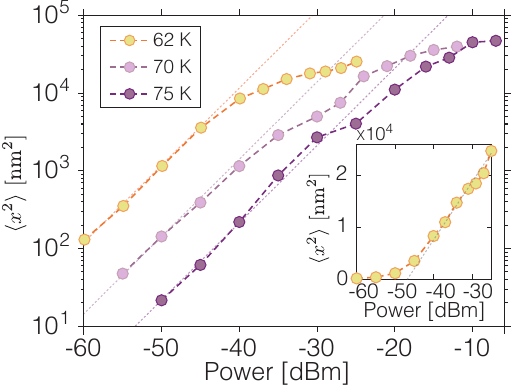}
        \caption{\textbf{Nonlinear dissipation at high temperature.} 
        Driven mechanical fluctuations with increasing drive powers at various temperatures. Dotted lines represent the linear dependence at low drive powers.
        Inset: linear-log plot for data at \SI{62}{\kelvin}. The grey dotted line represents a logarithmic dependence.}
        \label{fig:6}
    \end{figure}
    %%%%%%%%%%%%%%%%%%%%%%%%%%%%%%%%%%%%%%%%%%%%%%%%%%%%%%%%%%%%%%%%%%

    \subsection{Vortex dynamics at high temperature}
    At higher temperatures, vortices are activated by the oscillating field to overcome pinning barriers and creep between pinning sites rather than being strongly locally pinned. This process can be described by the Kim-Anderson flux-creep picture~\cite{anderson_1964_hard_RMP,blatter_1994_vortices_RMP}, in which the dissipation exhibits an Arrhenius-type exponential dependence on excitation energy as $\exp({-U(j)/k_{\mathrm{B}}T})$, where $U(j)=U_0(1-j/j_{\mathrm{c}})$ is the barrier energy dependent on the mechanical motion induced surface currents $j$ that is proportional to the motion velocity, and the critical current density $j_{\mathrm{c}}$.

    Figure~\ref{fig:6} shows driven mechanical fluctuations as a function of the power of an on-resonance drive at several temperatures. It is evident that at lower drives, the excited motion is proportional to the drive power, whereas at higher drive, we observe a slow logarithmic increase of the oscillation amplitude with increasing drive power, consistent with the onset of exponential dissipation from vortex creeping~\cite{grosser_1995_damping_APL, grosser_2000_vortex_JLTP, druge_2014_damping_NJP}. The effect of vortex creeping is further justified by the non-reversible mechanical frequency shift after cooling down the highly-excited magnet back to \SI{6}{\kelvin} as a result of vortex reconfiguration. 
    
%%%%%%%%%%%%%%%%%%%%%%%%%%%%%%%%%%%%%%%%%%%%%%%%%%%%%%%%Outlook%%%%%%%%%%%%%%%%%%%%%%%%%%%%%%%%%%%%%%%%%%%%%%%%%%%%%%
\section{Discussion and outlook} 
    Our work unravels an important loss mechanism of Meissner levitated systems. Even with $\operatorname{type-I}$ superconductors, vortices are inevitably present at defects~\cite{hofer_2023_high_PRL,gutierrez_2023_superconducting_PRApplied,smit_2025_superconducting_arXiv}, limiting mechanical quality factors. The model described in this work quantitatively estimates the damping associated with individual vortices. Although there are remaining puzzles to fully explain our observations, this work points out a path towards realizing highly coherent levitated systems, by either materials improvement~\cite{bland_2025_millisecond_nature}, or engineering artificial defects~\cite{eley_2016_decoupling_SST}.

    Our observations further demonstrate that the micrometer-scale levitated magnet can serve as a unique local probe of superconductors' properties.
    Wide tunability of mode frequencies, motional directions, and excited energies expands the sensor’s operational flexibility and facilitates the extraction of richer information. 
    Unlike other local near-field techniques that are mainly susceptible to surface effects~\cite{hoffman_2002_imaging_science} or conventional magnetic force microscopies that are primarily sensitive to static force~\cite{chen_2007_two_natphy}, this magnet probes vortices interacting with pinning centers deep within the bulk, and allows the study of vortex-induced dissipation~\cite{auslaender_2009_mechanics_natphy,eley_2017_universal_natmat,hovhannisyan_2025_scanning_Comm_Materials,buchacek_2019_experimental_PRB}. Our observations of telegraph signals in dynamics of levitated magnets indicate that we are able to resolve influences of individual vortices~\cite{nsanzineza_2014_trapping_PRL}. Accessing such pinning properties of a single vortex is essential for testing theories of superconductivity and improving the performance of superconducting materials. 
    In particular, decoherence of superconducting qubits is associated with the motion of vortices~\cite{bahrami_2025_vortex_arXiv}. Vortex configuration can induce fluctuations in qubit parameters, thereby limiting the performance and scalability of large-scale quantum processors~\cite{nsanzineza_2014_trapping_PRL,nambisan_2025_quantum_arXiv}. Beyond dissipation and fluctuations, however, the presence of vortices can also suppress quasiparticle density, leading to a surprising improvement in qubit coherence times~\cite{wang_2014_measurement_NatureComm,nsanzineza_2014_trapping_PRL}. In addition, vortex depinning processes have been associated with the tuning effect in superconducting qubits~\cite{kennedy_2025_josephson_PRL}.
    The micrometer-scale, second-long coherence time makes the levitated magnet an important, sensitive characterizing probe for superconducting devices, enabling studies of vortex-induced loss, tackling the microscopic sources of fluctuators, and investigating the interaction of vortices with the environment.

    We note that in the present system, multiple vortices interact with the magnet, each experiencing a distinct and complex pinning potential. While the current measurements are insufficient to directly isolate individual vortices, this should be possible, for instance, by resolving each vortex using imaging methods~\cite{wells_2015_analysis_SciRep, schlussel_2018_wide_PRApp}, manipulating individual vortices~\cite{straver_2008_controlled_APL}, or by reducing the number of trapped vortices, which is an important direction for future work.

    Although vortices are typically treated as semi-classical objects, their quantum behavior has been proposed theoretically ~\cite{olson_2004_vortices_PRL}, and observed experimentally~\cite{wallraff_2003_quantum_nature,fruchter_1991_low_PRB,Dutta_2021_Evidence_PRB}. Recent work has revealed coherence in superconducting vortex states~\cite{nambisan_2025_quantum_arXiv}, where a single vortex behaves as an effective two-level quantum system with modest coherence time and coherent controls. The strong coupling between such vortex qubits and a levitated magnet makes this a promising platform for realizing macroscopic quantum states~\cite{roda_2024_macroscopic_PRL}, potentially allowing one to test collapse models~\cite{bassi_2013_models_RMP}, providing insight into the interplay between quantum mechanics and gravity~\cite{belenchia_2018_quantum_PRD,bose_2017_spin_PRL}, and enabling novel quantum-enhanced sensing capabilities~\cite{munro_2002_weak_PRA}.
%%%%%%%%%%%%%%%%%%%%%%%%%%%%%%%%%%%%%%%%%%%%%%%%%%%%%%%%%%%%%%%%%%%%%%%%%%%%%%%%%%%%%%%%%%%%%%%%%%%%%%%%%%%%%%%%%%%%%%%%%%%%%%%%%%%%%%%

\begin{acknowledgments}
    We thank V. Geshkenbein, B. Stickler, J. G. E. Harris, E. Demler, J. Hoffman, S. Eley,  and S. Chattopadhyay for useful discussions, A. Cui and S. Lim for assistance with magnet magnetization, C. M. Brooks and Z. Hasan for assistance with XRD measurements, A. Jiang for assistance with PPMS measurements, J. G. E. Harris and B. Stickler for comments on the manuscript, and J. MacArthur for technical assistance. 
    MPMS measurements were performed at the Laukien-Purcell instrument center, a part of Harvard University.
    This work was supported by 
    the NSF Center for Ultracold Atoms, 
    Amazon Web Services (grant No.~A60290), 
    NSF (grant No.~OMA-2121044),
    DOE Quantum Systems Accelerator Center (grant No.~DE-AC02-05CH11231), 
    and the Air Force Office of Scientific Research (grant No.~FA9550-23-1-0333). 
    Y. W. acknowledges support from the HQI Postdoctoral Fellowship Program. 
    T. M. acknowledges support from the NSF Graduate Research Fellowship Program (grant No.~2140743). \\
\end{acknowledgments}
%%%%%%%%%%%%%%%%%%%%%%%%%%%%%%%%%%%%%%%%%%%%%%%%%%%%%%%%%%%%%%%%%%%%

\noindent
\textbf{Author contributions}:
M.L. and Y.W. conceived of the presented study. Y.W. and T.M. wrote the original draft. T.M., Y.W., J.S. and A.N. performed the experiments. Y.W. performed the theoretical modelling and the numerical simulations. Y.W., T.M, and J.S. analyzed the data. M.L supervised the work. All authors contributed to reviewing and editing the manuscript.\\

%%%%%%%%%%%%%%%%%%%%%%%%%%%%%%%%%%%%%%%%%%%%%%%%%%%%%%%%%%%%%%%%%%%%

\noindent
\textbf{Data availability}:
{All data that support the plots within this paper and other findings of this study are available from the corresponding author upon reasonable request.}

%%%%%%%%%%%%%%%%%%%%%%%%%%%%%%%%%%%%%%%%%%%%%%%%%%%%%%%%%%%%%%%%%%%%

\renewcommand{\appendixname}{APPENDIX}

\appendix

\newcommand{\expect}[1]{\langle {#1} \rangle}
\newcommand{\abs}[1]{\vert {#1} \vert}

%%%%%%%%%%%%%%%%%%%%%%%%%%%%%%%%%%%%%%%%%%%%%%%%%%%%%%%%%%%%%%%%%%%%
\begin{figure}
    \centering
    \includegraphics[width=1\linewidth]{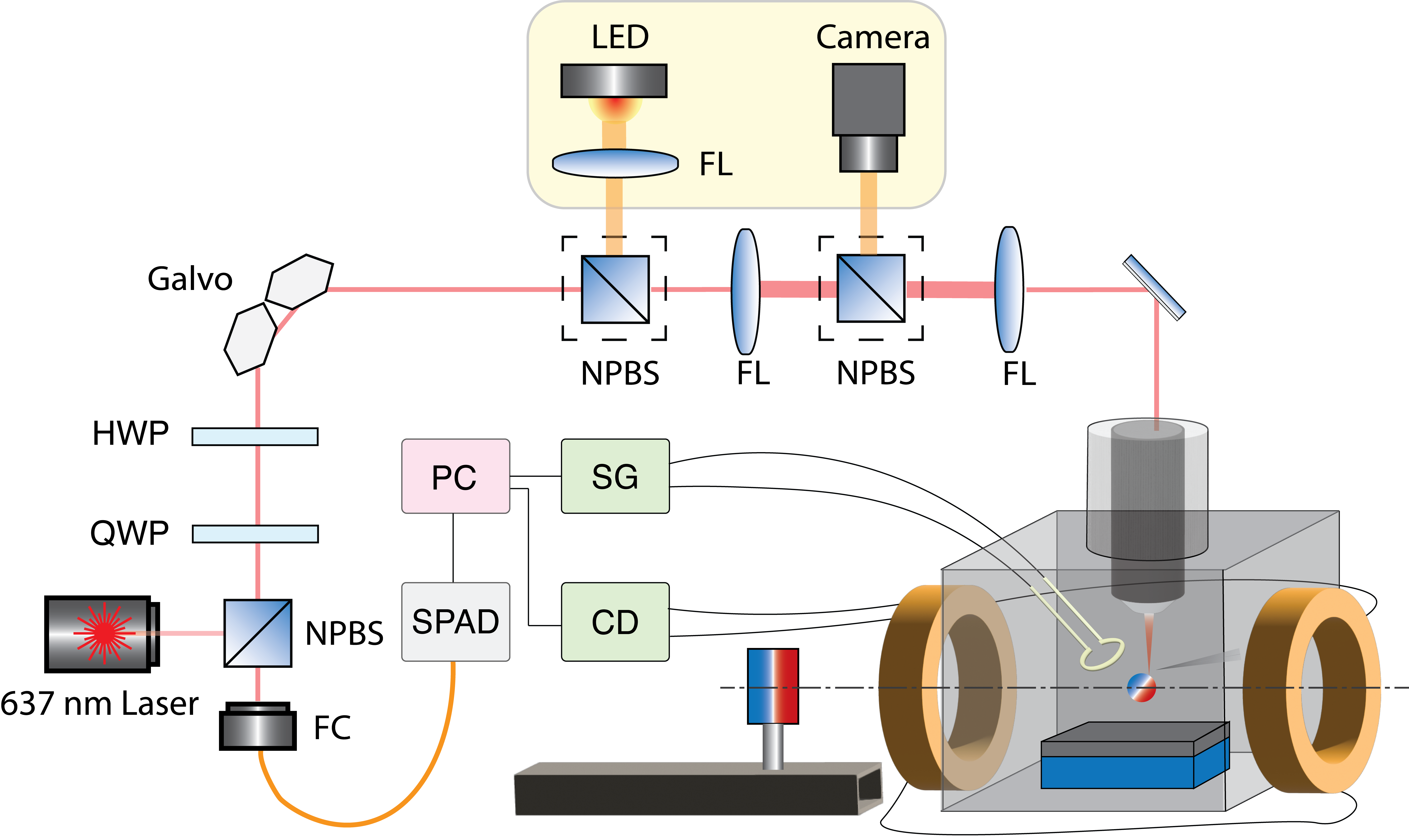}
    \caption{\textbf{Schematic of experiment setup. }The YBCO sample and an \textit{in-situ} micro-manipulation setup are placed in a cryostat integrated with a 4$f$ confocal microscope. The galvanometer allows for 2D scanning of a \SI{637}{\nano\meter} laser used for mechanical readout. The beam polarization is changed by quarter- and half-wave plates. The reflected beam passes through a beamsplitter, coupled into a fiber, and detected using an SPAD. The LED and camera used for coarse imaging are introduced using beamsplitters when necessary. A permanent magnet and a pair of Helmholtz coils outside the cryostat are used to tune the mechanical modes. NPBS: non-polarized beamsplitter; HWP/QWP: half/quarter waveplate; FC: fiber coupler; Galvo: Galvanometer; FL: focal lens; SG: Signal generator; CD: current driver.}
    \label{Appen_fig_1}
\end{figure}
%%%%%%%%%%%%%%%%%%%%%%%%%%%%%%%%%%%%%%%%%%%%%%%%%%%%%%%%%%%%%%%%%%

\section{EXPERIMENTAL SETUP AND CHARACTERIZATIONS}
    \subsection{Optical setup}
        \label{sec:Experiment setup}
        A diagram of our experiment optical setup is shown in Fig.~\ref{Appen_fig_1}. We use a \SI{637}{\nano\meter} laser whose polarization can be tuned using a quarter-wave plate and a half-wave plate to maximize the signal. The beam is steered using a galvanometer through a 4$f$ confocal microscope and focused with a $100\times$ objective with NA=0.8 (100$\times$ Nikon CFI60 TU Plan Epi ELWD Infinity Corrected Objective) inside the cryostat, allowing in-plane 2D scans of the laser focal point. The light reflected off the particle passes back through the same optical path, then is measured with a single photon avalanche detector (SPAD), where the photon count can be recorded as a function of arrival time. We use a camera during the levitation procedure and for coarse characterization. The broad-spectrum LED and camera paths are introduced into the main path using two beamsplitters (black dashed boxes). Both beamsplitters are flipped away from the optical path during mechanical measurements. A gold loop (\SI{25}{\micro\meter} diameter) next to the magnet can drive the motion of the particle via oscillating magnetic fields, which allows us to excite mechanical motion for ringdown measurements. 

    %%%%%%%%%%%%%%%%%%%%%%%%%%%%%%%%%%%%%%%%%%%%%%%%%%%%%%%%%%%%%%%%%%
    \begin{figure}
        \centering
        \includegraphics[width=0.5\linewidth]{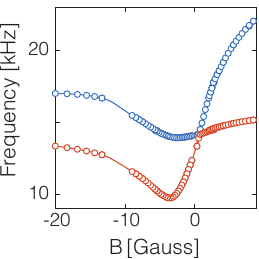}
        \caption{\textbf{Magnetic field sweep. }Frequencies of two in-plane translational modes during a $B$ field sweep via movement of a permanent magnet.}
        \label{Appen_fig_2}
    \end{figure}
    %%%%%%%%%%%%%%%%%%%%%%%%%%%%%%%%%%%%%%%%%%%%%%%%%%%%%%%%%%%%%%%%%%

    \subsection{Levitation protocol}
        \label{sec:Levitation protocol}
        During levitation, we micromanipulate the particle using a tapered tungsten needle attached to a 3-axis attocube stage. Above the critical temperature of the superconductor, a \SI{1.5\pm0.1}{\micro\meter} radius spherical magnet is picked up and positioned \SI{2.9\pm0.3}{\micro\meter} above the superconductor through surface forces between the needle and the particle. The cryostat is then cooled below the critical temperature such that the magnetic trap is formed. The restoring forces from the trap now overcome surface forces between the particle and the needle, allowing the needle to be moved away while the particle remains levitating in place. 
          
    %%%%%%%%%%%%%%%%%%%%%%%%%%%%%%%%%%%%%%%%%%%%%%%%%%%%%%%%%%%%%%%%%%
    \begin{figure*}
        \centering
        \includegraphics[width=0.9\linewidth]{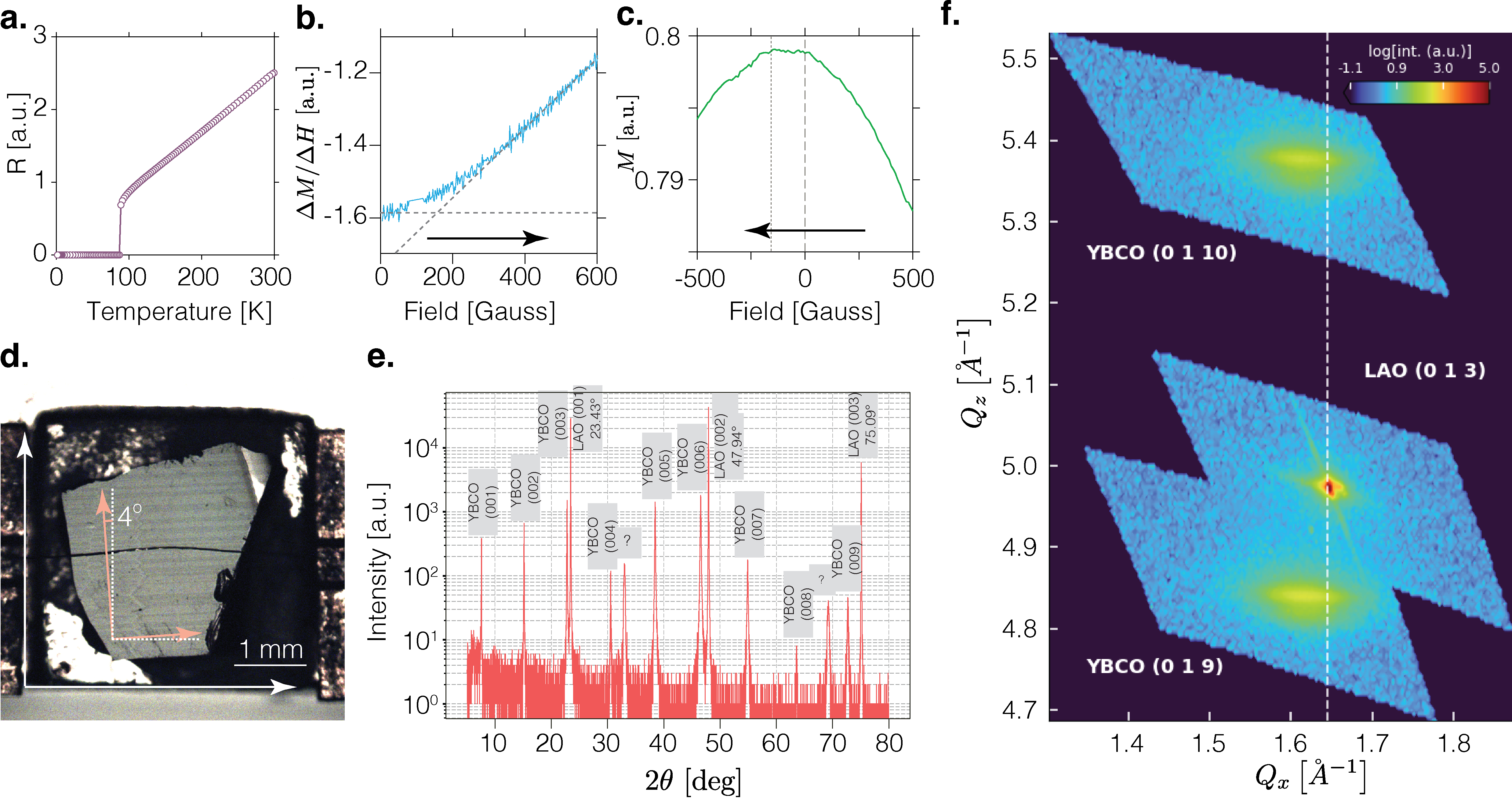}
        \caption{\textbf{Characterization of YBCO sample. }(a) Resistance as measured through a four-wire measurement as a function of temperature. A sharp drop in the resistance is seen at \SI{87\pm1}{\kelvin}, marking the material's critical temperature. The normal state resistance of YBCO is approximately proportional to its temperature. (b) Differential magnetic susceptibility as the applied field is first increased from zero to \SI{1.5}{\tesla}. Dashed grey lines indicate the initial constant susceptibility and later linearly increasing susceptibility with a crossover at approximately \SI{157}{G}. (c) Magnetization of YBCO as the field is decreased from \SI{1.5}{\tesla} through zero and towards \SI{-1.5}{\tesla}. Dashed vertical lines indicate zero field and \SI{-157}{G}. (d) Optical image of the YBCO sample taken under 2.5$\times$ magnification. Sample holder edges, aligned to the lab frame for angle measurements, are indicated in white arrows. The LAO in-plane lattice vectors are indicated in pink arrows.
        (e) XRD $\operatorname{\theta -2\theta}$ scan of YBCO. Families of peaks corresponding to LAO (0\,0\,1) and YBCO (0\,0\,1) and their higher orders are labeled. Two unidentified peaks are indicated by question marks. (f) Reciprocal space map of the YBCO and LAO, showing peaks associated with the (0\,1\,10) and (0\,1\,9) reciprocal lattice points of YBCO and (0\,1\,3) of LAO. The dashed line highlights the mean LAO reciprocal lattice point along the $x$ reciprocal direction, mostly overlapping with that of the film.}
        \label{Appen_fig_3}
    \end{figure*}
    %%%%%%%%%%%%%%%%%%%%%%%%%%%%%%%%%%%%%%%%%%%%%%%%%%%%%%%%%%%%%%%%%%

    \subsection{Magnetic field $B$ sweep}
        \label{sec:b_sweep}
        The magnetic field applies a torque to the levitated micromagnet, rotating its magnetization axis as well as its corresponding image dipole, enabling tuning of resonance frequencies for various experiments. We employ two magnetic field sources: a NdFeB permanent magnet on a micrometer translational stage for a coarse sweeping and a pair of Helmholtz coils in proximity to the sample for a fine tuning. Each system produces an approximately homogenous magnetic field at the site of the micromagnet in a direction parallel to the $x$ mode. Figure~\ref{Appen_fig_2} shows resonance frequencies of $x$ and $y$ modes across a coarse sweep of $\sim 30$ G. Both mode frequencies plateau around \SI{-20}{G} due to the magnetization axis of the particle aligning with the applied magnetic field. With a fixed permanent magnet position, a fine sweep with a range around \SI{150}{\micro\tesla} and a resolution of \SI{30}{\nano\tesla} using the Helmholtz coils is shown in Fig.~1.

    \subsection{YBCO characterization}
        \label{sec:YBCO characterization}
        We procure YBCO films from 2D Semiconductors. \SI{270}{\nano\meter} of YBCO are PLD deposited onto both sides of a 2-inch \SI{0.5}{\milli\meter} thick wafer of mechanically polished single crystal LAO and subsequently cleaved into 2 mm pieces used in the experiment. We characterize the YBCO film using resistive, magnetic moment, and X-ray measurements.  
        \begin{description}
             \item[Resistance]{Pieces of the YBCO film are wirebonded in a four-point configuration and cooled to \SI{1.8}{\kelvin} in a PPMS chamber (PPMS, Quantum Design). We measure resistance as a function of temperature, finding the $T_{\mathrm{c}}$ of these films to be \SI{87(1)}{\kelvin}. We extract the RRR ($R(300~\text{K})/R(100~\text{K})$) from this curve to be 3.67. The resistance of the normal conductor part is nearly proportional to the temperature.
             }
            
            \item[MPMS]{We measure the magnetization hysteresis curve of the YBCO films using a SQUID magnetometer (MPMS, Quantum Design). Two millimeter square samples of YBCO are cooled to \SI{4}{\kelvin} in a zero magnetic field environment. The magnetic moment of the sample is then measured as a magnetic field is applied perpendicular to the sample. We increase the magnetic field from 0 to \SI{1.5}{\tesla}, then decrease the field back to zero. We then apply a field in the opposite direction from 0 to \SI{1.5}{\tesla} and decrease the field back to zero. When first increasing the field from zero, we find the magnetization susceptibility deviates from a constant value around \SI{157}{G}, shown in Fig.~\ref{Appen_fig_3}(c), suggesting the field has surpassed the material's $H_{c1}$. Similarly, after ramping the field and changing the applied field direction, we see the magnetization deviate from a constant value at roughly \SI{-157}{G}.
            }
            
            \item[XRD]{We perform standard X-ray diffraction (XRD) $\operatorname{\theta-2\theta}$ measurements of the YBCO film that is used in the experiment to obtain information about its phase and crystallographic orientation, shown in Fig.~\ref{Appen_fig_3}(e). We find families of peaks primarily corresponding to (0\,0\,1) LAO and (0\,0\,1) YBCO and their corresponding higher orders, indicating $c$-axis of the YBCO and the $c$-axis of the substrate LAO are aligned along the z-axis of the lab frame as we expect. 
            
            We also perform $\operatorname{\omega-2\theta}$ measurements to obtain reciprocal space maps (RSMs) of both the film and the substrate, as shown in Fig.~\ref{Appen_fig_3}(f). Peaks corresponding to the (0\,1\,9) and (0\,1\,10) reciprocal lattice points of YBCO are mostly aligned with the (0\,1\,3) reciprocal lattice point of LAO along the x-direction (white dashed line), indicating that the in-plane lattice orientation of YBCO is mostly aligned with that of the substrate. However, instead of two well-resolved peaks associated with the $a$- and $b$- axes, we observe an elongated intensity profile along $Q_x$. We attribute this elliptical shape to several possible effects: (i) strain in the film that matches the symmetric in-plane lattice constant of LAO at the interface and relaxes away from the substrate, (ii) a mixture of $a\operatorname{-}b$ and $b\operatorname{-}a$ stacking relative to the substrate, (iii) excess structural or chemical disorder within the film, or (iv) measurement uncertainty arising from the small sample volume. To further test the origins of this profile, we rotated the sample by $90^{\circ}$ and repeated the same $\operatorname{\omega-2\theta}$ measurement. The RSM exhibited an almost identical reciprocal-lattice profile, suggesting that the YBCO film lacks a well-defined global in-plane anisotropy.
            
            From these measurements, we are able to determine the orientation of the LAO in-plane lattice (pink arrows), which is aligned with YBCO, with respect to the sample. The in-plane lattice orientation is also aligned with the straight cleaved edge of the substrate as we expect. By comparing microscopic images of the sample in the XRD measurement and the levitation experiment, we estimate the LAO in-plane lattice is $+4(1)^{\circ}$ tilted from the lab frame (white arrows) that we used to determine the motional directions in Fig.~3 of the main text, as shown in Fig.~\ref{Appen_fig_3}(d). 
            }
        \end{description}

\section{SYSTEM MODELING AND ESTIMATION}
    \subsection{Displacement sensitivity estimation}
        \label{sec:Displacement sensitivity estimation}
\begin{figure}
        \centering
        \includegraphics[width=1\linewidth]{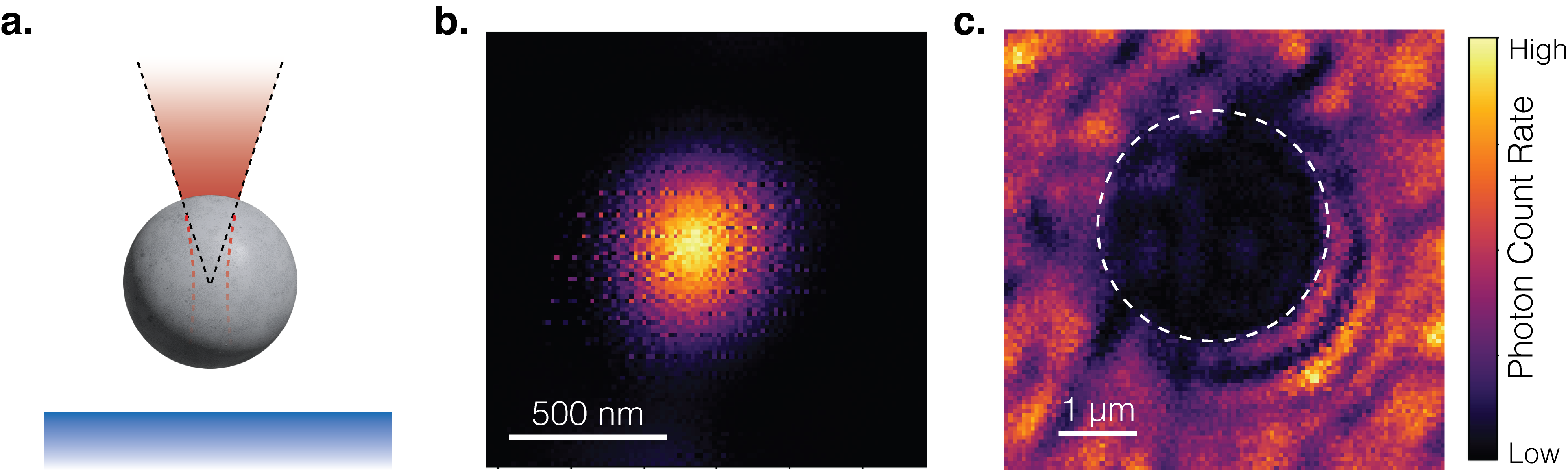}
        \caption{\textbf{Optical Readout.}
        (a) A schematic of the laser focused on the spherical magnet. The red dashed lines represent the Gaussian beam profile, and the black dashed lines indicate its asymptotic behavior. When the laser is focused near the center of the magnet, the incident beam is nearly perpendicular to the surface, thereby maximizing the reflected light count rate. 
        (b) An example 2D galvo scan when count rate is maximized. The color of each pixel represents the count rate of the reflected light when the laser is steered to the corresponding location. (c) A 2D galvo scan when the laser is focused on the YBCO surface. The white dashed circle highlights the shadow created by the levitating magnet, which is used to estimate the size of the magnet.}
        \label{Appen_fig_4}
    \end{figure}

        We maximize the count rate of light reflected off the particle by focusing the laser near the center of the spherical magnet, as shown in Fig.~\ref{Appen_fig_4}(a). A 2D spatial galvo scan in this case reveals the expected Gaussian beam profile for a highly focused beam, shown in Fig.~\ref{Appen_fig_4}(b). This profile can be fit to the Gaussian function $N_0e^{-x^2/2w_0^2}$ yielding a full-width-half-maximum \SI{494}{\nano\meter}, which is close to the Airy diffraction limit (\SI{486}{\nano\meter}) of \SI{637}{\nano\meter} wavelength laser through an NA=0.8 objective. The intensity derivative corresponds to the relative displacement response, which is maximized when the center of the laser is focused at $x_0=w_0$. More specifically, the measurement sensitivity is given by
        \begin{align}
        \label{eq:measurement_sensitivity}
        S_{xx}=S_{NN}\abs{\frac{\partial x}{\partial N}}^2={\bar{N}}\abs{\frac{\partial x}{\partial N}}^2\geq\frac{1.4\times10^{7}}{\bar{N}},
        \end{align}
        where $\bar{N}$ is the average count rate, $S_{xx}$ is the noise spectrum of the displacement $x$, and $S_{NN}$ is noise spectrum of the count rate $N$,  limited by shot noise as $S_{NN}=\bar{N}$. Assuming we have collected a photon count rate of 1~Mcps (limited by saturation count rate of SPAD), we achieve a displacement sensitivity of \SI{0.4}{\nano\meter\per\sqrt{\hertz}}. 
         A related discussion can be found in Ref.~\cite{hansen_2025_optical_arXiv}.

    %%%%%%%%%%%%%%%%%%%%%%%%%%%%%%%%%%%%%%%%%%%%%%%%%%%%%%%%%%%%%%%%%%

    %%%%%%%%%%%%%%%%%%%%%%%%%%%%%%%%%%%%%%%%%%%%%%%%%%%%%%%%%%%%%%%%%%
        
    \subsection{Linear coupled mode theory}
        \label{sec: linear_coupling}
        A linear coupled mode theory is used to describe the avoided crossing feature between $x$ and $y$ modes in Fig.~\ref{fig:3}(b), which allows setting nearly all directions of motion by sweeping $B$ within \SI{1}{G}. 
        Assuming uncoupled frequencies $\omega_x(B)$ and $\omega_y(B)$ of $x$ and $y$ modes linearly depend on $B$ and a constant coupling rate $g_1$ between each other, the coupled mode equation is
        \begin{align}\label{eq:CML_eq}
            \left[\begin{array}{cc}
                \omega_x(B) & g_1 \\
                g_1 & \omega_y(B) 
            \end{array}\right]
            \left[\begin{array}{l}
                x \\ y
            \end{array}\right]=\omega
            \left[\begin{array}{l}
                x \\ y
            \end{array}\right],
        \end{align}
        where $\omega$ are the eigenfrequencies of hybridized modes. The motional directions of normal modes change as $x$ and $y$ modes get hybridized. The angle throughout the avoided crossing can be measured using galvanometer scans when modes are driven. We measure the absolute motional angle $\theta$ with respect to our galvanometer scan axes (lab frame). The motional angle $\theta$ can be solved in Eq.~\eqref{eq:CML_eq} as
        \begin{align}\label{eq:mixed_angle}
          \theta(B)=\frac{1}{2}\tan^{-1}(\dfrac{2g_1}{\omega_x(B)-\omega_y(B)}).
        \end{align}
        A global fitting to this model is shown in Fig.~\ref{fig:3}(c,d), yielding $g_1/2\pi=$ \SI{84.2\pm 0.2}{\hertz} and the center of the avoided crossing at \SI{116.2\pm 0.4}{\micro\tesla}.

    \subsection{Estimation of number of vortices}
        \label{sec:Estimation_of_number_vortices}
\begin{figure}
        \centering
        \includegraphics[width=1\linewidth]{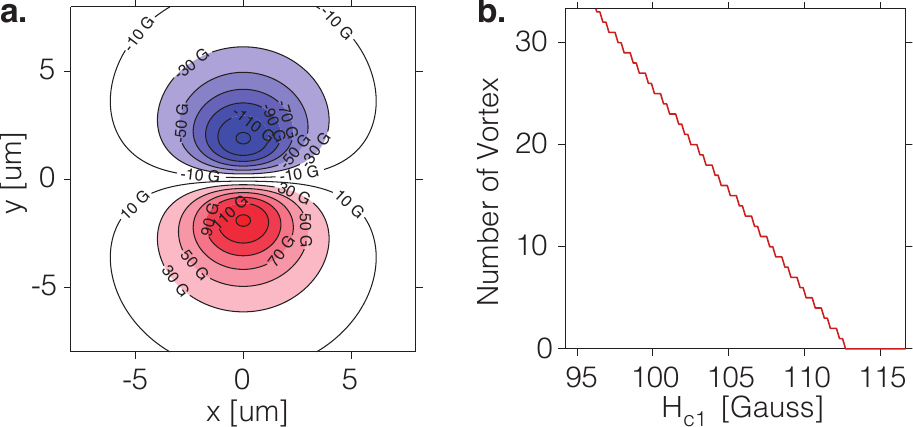}
        \caption{\textbf{Estimating the number of trapped vortices. }
        (a) Magnetic field distribution at the surface of YBCO. Field lines are estimated assuming a magnet size of $r=$\SI{1.5(1)}{\micro\meter}, levitation height of $h=$\SI{3.0(1)}{\micro\meter}, magnetic remanence of \SI{0.8}{\tesla}, and horizontal magnetization orientation. 
        (b) Calculated total number of vortices as a function of the lower critical field $H_{c1}$ for the referenced levitation parameters.}
        \label{Appen_fig_5}
    \end{figure}

        The minimal unit of the magnetic flux in a superconductor is a vortex, whose magnetic flux is $\Phi_0={h}/{2e}=2.07\times10^{-15} ~\mathrm{Wb}$. The upper bound of the number of vortices from the magnet can be estimated by
        ${\Phi_{\text{total}}}/{\Phi_0}={\int_{\mathcal{S}}\vert B_z\vert\mathrm{d}x\mathrm{d}y}/{\Phi_0}$
        where $\mathcal{S}$ is the top surface of the superconductor and $z$ is the $\mathcal{S}$ surface normal. If we assume all fields below the lower critical field  $H_{c1}$ will be repelled from the material during the field-cooling process, the number of vortices is
        \begin{align}
            \frac{\Phi_{\text{total}}}{\Phi_0}=\frac{\int_{\mathcal{S}(B_z>H_{c1})}\vert B_z\vert\mathrm{d}x\mathrm{d}y}{\Phi_0},
        \end{align}
        where $\mathcal{S}(B_z>H_{c1})$ represents the surface where $B_z$ is greater than $H_{c1}$. In reality, this number depends on properties of pinning centers, the exact field-cooling process, material properties, \textit{etc}. For a magnet of radius $r=$\SI{1.5(1)}{\micro\meter} with a levitation height $h=$\SI{3.0(1)}{\micro\meter}, magnetic remanence $B_r=$\SI{0.8}{\tesla}, and a horizontal magnetization orientation, the corresponding magnetic field distribution and the number of vortices as a function of the lower critical field $H_{c1}$ are shown in  Fig.~\ref{Appen_fig_5}.
         
        Alternatively, the magnetic field of a single vortex can be described by a monopole-monopole model~\cite{auslaender_2009_mechanics_natphy}. If the penetration depth is negligible, we have~\cite{chang_1992_scanning_APL}
        \begin{align}
            \vec{B}(\vec{r}, z) \approx\frac{\Phi_0}{2 \pi} \frac{\left(\vec{r}+\vec{z}\right)}{\left(r^2+z^2\right)^{3 / 2}},
        \end{align}
        where $\vec{r}$ is the magnet's in-plane position from the vortex and $z$ is the magnet's vertical distance from the surface. Hence, the effective number of vortices can be estimated as
        \begin{align}\label{eq:single_vortex_spring}
            \frac{k}{k_\text{m}}\approx\frac{m\omega_{\mathrm{m}}^2\pi z^4}{3\Phi_0\vec{M}\cdot\hat{z}}\approx\frac{\mu_0\rho\omega_{\mathrm{m}}^2\pi z^4}{3\Phi_0B_{\mathrm{r}}},
        \end{align}
        where $k=m\omega_{\mathrm{m}}^2$ is the effective spring constant of the mechanical mode, $k_{\text{m}}$ is the spring constant provided by a single vortex, $B_{\mathrm{r}}$ is the magnet remanence, $\rho$ is the magnet density, and $\mu_0$ is the vacuum permeability. Here, we simplify the estimation by assuming all vortices are directly under the magnet ($\vec{r}=0$) and the magnetic moment is vertical. Based on this model, the effective number of vortices in our case is less than 4. In reality, the spatial distribution of trapped vortices should be accounted for a more accurate estimation.

    %%%%%%%%%%%%%%%%%%%%%%%%%%%%%%%%%%%%%%%%%%%%%%%%%%%%%%%%%%%%%%%%%%
    %%%%%%%%%%%%%%%%%%%%%%%%%%%%%%%%%%%%%%%%%%%%%%%%%%%%%%%%%%%%%%%%%%

     \subsection{Eddy current damping estimation}
        \label{sec:eddy_current_estimate}
        Eddy current damping is a common dissipation source of an oscillating magnet. Several levitation experiments argue that this is the dominating loss mechanism in their setups~\cite{gutierrez_2023_superconducting_PRApplied,timberlake_2024_linear_PRR}. In this section, we briefly discuss the possibility of  eddy currents damping being a limit of the mechanical quality factor in our system.
        
        Consider an ideal point dipole $\vec{M}$ (say $\vec{M}=M\hat{x}$) lying parallel to an infinite, thin, isotropic conducting plate of conductivity $\sigma$. For a translation with velocity $\vec{v}$, the conductor experiences an electric field $\vec{E}=-\vec{v}\times\vec{B}$. Thus, surface Joule heating is
        \begin{align}
            \Delta E =\int\int_{\mathcal{S}} \sigma \abs{\vec{E}}^2\mathrm{d}\mathcal{S}\mathrm{d}t=\int\int_{\mathcal{S}} \sigma \abs{\vec{v}}^2B_z^2\mathrm{d}\mathcal{S}\mathrm{d}t,
        \end{align}
        where we only consider currents flowing parallel to the thin plate, and $B_z$ is the $B$ field perpendicular to the conductor surface. Assume the particle is oscillating at $\omega_{\mathrm{m}}$ with a small amplitude $a$, the energy loss per oscillation is
        \begin{align}
            \Delta E =\sigma \frac{\tau_0}{2}a^2\omega_{\text{m}}^2\int_{\mathcal{S}} B_z^2\mathrm{d}\mathcal{S},
        \end{align}
        where $\tau_0=2\pi/\omega_{\mathrm{m}}$ is the time per oscillation. Therefore, the quality factor is
        \begin{align}\label{eq:q_eddy_current}
            Q=\frac{E}{\Delta E}=\frac{\frac{1}{2}m\omega_{\text{m}}^2a^2}{\sigma \frac{\tau_0}{2}a^2\omega_{\text{m}}^2\int_{\mathcal{S}} B_z^2\mathrm{d}\mathcal{S}}=\frac{m\omega_{\mathrm{m}}}{\sigma 2\pi\int_{\mathcal{S}} B_z^2\mathrm{d}\mathcal{S}}.
        \end{align}
        Equation~\eqref{eq:q_eddy_current} suggests the quality factor of eddy current damping should be independent of in-plane motion directions and only proportional to the corresponding mode frequency. In particular, in our angle-dependent damping measurements, we achieve different motion directions via mode hybridization when $x$ and $y$ mode frequencies are nearly identical. Therefore, we would expect the associated quality factors to be nearly identical if the damping is dominated by eddy currents. This is the opposite of what we have observed in Fig.~3(b) of the main text. It is worth noting that the dissipation for the $z$ mode is different, because the eddy current damping depends on $B_{\parallel}$ rather than $B_{z}$.  
        
        Multiple experimental observations suggest that the dissipation in our system is unlikely to be limited by eddy current damping. The closest surface to the levitated magnet is the  $\operatorname{type-II}$ superconducting substrate, which suppresses changes in perpendicular magnetic flux that are essential for generating eddy currents. Consequently, the primary mechanism required for dissipative induction is strongly hindered. Second, to achieve the mode hybridization displayed in Fig.~\ref{fig:3} of the main text, an external magnetic field was applied to tune the in-plane translational mode frequencies so that the $x$ and $y$ modes became nearly degenerate. The emergence of this symmetry indicates that the magnetic moment of the levitated particle is oriented close to the $\hat{z}$ direction, thereby yielding geometric symmetry for in-plane motion. This configuration eliminates magnetization-orientation-based asymmetries that potentially lead to anisotropic dissipation. Third, we observe that the mechanical dissipation increases sharply with increasing temperature of the YBCO substrate, whereas the electrical conductivity of the normal (non-superconducting) component of the sample decreases with temperature, as shown in Fig.~\ref{Appen_fig_3}(a). 
        Thus, eddy current dissipation is expected to decrease with increasing temperature, which is inconsistent with the trend measured in Fig.~\ref{Appen_fig_7}. Finally, additional experimental features, including strong magnetic-field dependence of damping (Appendix~\ref{sec:Temperature dependence at different B}), abrupt dissipation jumps (Fig.~\ref{fig:2} of the main text), and nonlinear damping behavior (Fig.~\ref{fig:4} of the main text), differ qualitatively from conventional eddy current damping. Taken together, these observations strongly suggest that eddy current mechanisms cannot account for the dominant source of dissipation in our system.

    \subsection{Gaussian kernel density estimator}
        \label{sec:Gaussian_kernel_density_estimator}
        The probability density function (PDF) of the inverse quality factor $1/Q$ is obtained by the Gaussian kernel density estimator. Here, we fitted each ringdown with a single exponential function $y=ae^{-2\pi ft/Q}+b$, where $y$ is the spectrum area around the resonance frequency, $f$ is the frequency of the mode, $Q$ is the quality factor, and $a,b$ are fitting parameters. To remove the bias from nonlinear fitting methods, we linearize the fitting function to $\log(y-b)=\log a-2\pi f/Q t$, where $b$ is directly estimated by the mean value of the thermal fluctuation at the end of each ringdown. If we assume the fluctuation of $\log(y-b)$ is random, independent, and fixed in a single ringdown, the estimator of the inverse quality factor follows
        \begin{align}
            \hat{1/Q}\sim \mathcal{N}(1/Q_i,\sigma_i),
        \end{align}
        where values of $1/Q_i,\sigma_i$ are extracted from the ordinary least-squares estimation for the \textit{i}-th ringdown. Repeated ringdown experiments generate a set of $\{1/Q_i\}$ and the corresponding $\{\sigma_i\}$. It is worth noticing that across different ringdown experiments, $1/Q_i$ is independently sampled, and the corresponding deviation $\sigma_i$ can be different. Thus, the PDF of $1/Q$ can be estimated by the Gaussian kernel estimator as 
        \begin{align}
            p(1/Q)=\frac{1}{N}\sum_{i=1}^{N}\mathcal{N}(1/Q_i,\sigma_i),
        \end{align}
        where each ringdown is weighted equally.     
        
\section{MECHANICAL-VORTEX MODEL}
    \label{sec:Mechanical-vortex interaction}

    \subsection{Theory of mechanical-vortex model}
    \label{sec:hybrid_mech_vortex_system}
        The levitated magnet above a $\operatorname{type-II}$ superconductor can be primarily explained by the frozen-image-dipole model~\cite{kordyuk_1998_magnetic_JAP,gieseler_2020_single_PRL}, which quantitatively describes the response of a mixed state superconductor to the displacement of the magnet. In this section, we discuss a correction to this model by considering the contribution of interactions between mechanical motion and vortices.
         
        The frozen-image-dipole model, which effectively applies Green’s function techniques, captures the boundary conditions of a type-II superconductor under a constant field from a magnetic dipole while cooling. The boundary magnetic field on the surface of the superconductor is treated as a combination of the field from a frozen dipole that generates the same magnetic field as the levitated magnet before cooling, the levitated magnet, and a corresponding image dipole to continually compensate for the field from the magnet. 
         In this model, after cooling the system below $T_{\mathrm{c}}$, the magnetic field distribution inside the superconductor is assumed to no longer respond to external magnetic field perturbations. This description is accurate only when the superconductor behaves as an absolutely hard superconductor, in which the magnetic field penetration is negligible. However, when levitating a micron-scale magnet a few micrometers above the superconductor, this model breaks down for two key reasons: 1. the penetration depth $\lambda$ is non-negligible compared to the levitation height; 2. only a few tens of vortices on the superconductor surface are involved in the levitation.
        
        A vortex interacts with randomly distributed pinning centers and meanders through the superconductor. The interplay between the pinning and vortex elasticity causes the vortex to behave rigidly on short length scales and flexibly on large length scales. Zooming in on one of the pinning sites, the pinning center acts like an anchor, constraining the vortex. However, this anchor is also associated with a pinning potential, in which the vortex may move back and forth. According to Ref.~\cite{doyle_1993_direct_PRL}, the restoring force is linear in a small displacement range ($<1~\text{nm}$) and becomes nonlinear beyond that range. Therefore, we can treat all pinning centers as harmonic potentials for the simplicity of the following discussion. 

    \subsection{Mechanical-vortex interaction}
        \label{SI_sec:Mechanical-vortex interaction}
        %%%%%%%%%%%%%%%%%%%%%%%%%%%%%%%%%%%%%%%%%%%%%%%%%%%%%%%%%%%%%%%%%
        \begin{figure}
            \centering
            \includegraphics[width=0.8\linewidth]{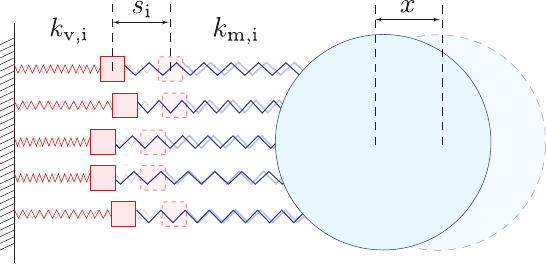}
            \caption{\textbf{Schematic of mechanical-vortex system. }The blue circle represents the magnet, and the red boxes represent the position of individual vortices in their pinning sites. Each blue spring represents 
            a single vortex line that originates from the magnet and ends in the YBCO pinning site. Red springs represent the effective harmonic potential of each pinning site.}
            \label{Appen_fig_6}
        \end{figure}
        %%%%%%%%%%%%%%%%%%%%%%%%%%%%%%%%%%%%%%%%%%%%%%%%%%%%%%%%%%%%%%%%%%

        The surface current generated by the oscillating magnet can be decomposed into $\vec{J}_{\text{Im}}$ and $\vec{J}_{\text{Fr}}$, corresponding to an image dipole and a frozen dipole, respectively. It is easy to show 
        \begin{subequations}
            \begin{align}
                \vec{B}_{\text{Im}}(\vec{r}, \vec{r}_{\text{M}})&=\vec{B}_{\text{M}}(\vec{r}-\vec{r}_{\text{M}},0),\\
                \vec{B}_{\text{Fr}}(\vec{r}, \vec{r}_{\text{M}})&=-\vec{B}_{\text{M}}(\vec{r},0),
            \end{align}
        \end{subequations}
        where $\vec{B}_{\text{M}}(\vec{r},0)$ is the in-plane magnetic field at $\vec{r}$ when the position of the magnet is at the origin $\vec{r}_M=0$, $\vec{B}_{\text{Im}}$ and $\vec{B}_{\text{Fr}}$ are respectively the in-plane magnetic fields from the frozen-dipole and image-dipole at $\vec{r}$ when the magnet is at $\vec{r}_{\mathrm{M}}$. The surface current $\vec{J_{\text{Im}}}$ corresponding to the image dipole is given by
        \begin{align}
            \vec{J}_{\text{Im}}(\vec{r})=2\hat{n}\times \vec{B}_{\text{Im}}(\vec{r})/\mu_0=2\hat{n}\times \vec{B}_{\text{M}}(\vec{r}-\vec{r}_{\text{M}})/\mu_0,
        \end{align}
        where $\hat{n}$ is the unit normal vector pointing outward from the top surface. Thus, the net surface current $\Delta \vec{J}$ due to the displacement of the magnet is
        \begin{align}
            \Delta \vec{J}(\vec{r})&=\vec{J}_{\text{Fr}}(\vec{r})+\vec{J}_{\text{Im}}(\vec{r})\nonumber\\
            &=
            -\vec{J}_{\text{M}}(\vec{r})+\vec{J}_{\text{M}}(\vec{r}-\vec{r}_{\text{M}})\nonumber\\
            &\approx - \vec{\nabla}\vec{J}_{\text{M}}(\vec{r})\cdot\vec{r}_{\text{M}},
        \end{align}
        where $\vec{J}_{\text{Fr}}(\vec{r})$ is the frozen dipole surface current and $\vec{J}_{\text{M}}(\vec{r})=2\hat{n}\times \vec{B}_{\text{M}}(\vec{r},0)/\mu_0$ is the surface current of the image dipole when $\vec{r}_{\text{M}}=0$. For a magnet oscillating with amplitude $a$, the magnet position is $\vec{r}_\text{M}=\hat{a}\vert a \vert \cos(\omega_{\mathrm{m}} t)$. For example, the displacement along the $x$-axis $\Delta x$ generates a net surface current
        \begin{align}\label{eq:surface_current}
            \Delta \vec{J}(\vec{r})=-\frac{2}{\mu_0}(\frac{\partial B_{\text{M},y}}{\partial x},-\frac{\partial B_{\text{M},x}}{\partial x})\Delta x.
        \end{align}
        The excited net surface current exerts a Lorentz force on each vortex, given by
        \begin{align}
            F_{v,i}=\Phi_0\hat{n}\times  \Delta \vec{J}(\vec{r}),
        \end{align}
        where $F_{\text{v,i}}$ is the Lorentz force on the \textit{i}-th vortex. As can be seen from Eq.~\eqref{eq:surface_current}, the Lorentz force is not necessarily along the direction of the motion. The net force from all vortices must align with the direction of the motion, as it comprises the net restoring force for the magnet. 

        The motion of the magnet will drive the vortices associated with the frozen dipole into motion. 
        The far-off-resonant vortex is always balanced by the restoring force of the pinning potential and the applied Lorentz force. Therefore, we have   
        \begin{align}\label{eq:adiabatic_balance}
            \Phi_0\Delta J+\frac{\partial V_\text{p,i}}{\partial s_{\mathrm{i}}}=0,
        \end{align}
        where $V_\text{p,i}$, $s_{\mathrm{i}}$ are the pinning potential and the displacement of the \textit{i}-th vortex, respectively. $V_\text{p,i}$ is approximately parabolic for $s\ll \xi$ as described before, so Eq.~\eqref{eq:adiabatic_balance} can be simplified as
        \begin{align}
            \Phi_0\frac{\partial J_0}{\partial x}(\Delta x-s_{\mathrm{i}})- k_{\text{v,i}}s_{\mathrm{i}}=0,
        \end{align}
        where we assume the displacement is along the $x$ direction without loss of generality, $k_{\text{v,i}}=-{\partial^2 V_{\text{p}}}/{\partial x^2}$ is the effective spring constant for \textit{i}-th vortex ($k_{\text{v,i}}=\alpha_{\text{i}}L$ where $L$ is the length of vortex and $\alpha_{\text{i}}$ is the Labusch parameter). 
        
        The motion of vortices creates an electric field $E=B_z\times \bar{v}_{\text{v}}= \sum_{i} v_{\text{v,i}} \Phi_0$, where $v_{\text{v,i}}=\alpha_{\text{i}} v_{\text{m}}$ is the velocity of the vortex, proportional to the magnet velocity $v_{\text{m}}$.
        Notice that the restoring force for the magnet should be identical to the Lorentz force on the vortex, giving $\alpha_{\text{i}}=k_{\text{m,i}}/(k_{\text{v,i}}+k_{\text{m,i}})$. As discussed in this Ref.~\cite{grosser_1997_detecting_arXiv}, the induced $E$ will produce ohmic losses in the normal core of the vortex, resulting in an energy dissipation rate of $W=\Delta J\cdot E$. For an oscillating magnet at frequency $\omega_\text{m}$ with an amplitude $a$, the dissipated energy per oscillation is
        \begin{align}
            \Delta E&=\int_0^\tau \int_{\mathcal{S}} \frac{\partial J_y}{\partial x} x \bar{v}_{\text{v}} B_z \mathrm{d}S \mathrm{d}t=2a^2\alpha \int_{S} \vert \dfrac{\partial J_y}{\partial x} B_z\vert \mathrm{d}S,
        \end{align}
        where $\tau=2\pi/\omega_{\mathrm{m}}$ is the period of one full oscillation. 
        Effectively, this can be rewritten as
        \begin{align}
            \Delta E_{\text{M}}=\frac{1}{2}R_{\text{s}}\tau \int_{\mathcal{S}}(\Delta J)^2\mathrm{d}S,
        \end{align}
        where the dissipation is extracted as ohmic losses by an effective surface resistance $R_{\text{s}}$.
    
        In a microscopic picture, the ohmic dissipation can be associated with the Bardeen-Stephen viscous force $f_{\text{vis}}=-L\eta_{\text{BS}}\dot{s}$~\cite{bardeen_1965_theory_PR}, where $\eta_{\text{BS}}={\Phi_0^2}/{2 \pi \xi^2 \rho_n}$ is the Barden-Stephen viscosity with $\rho_n$ being the resistivity of the electron fluid in the vortex core, often taken to be the normal state resistivity, $L$ is the length of the vortex, and $\xi$ is the coherence length. Therefore, the dissipation per oscillation per vortex is
        \begin{align}\label{eq:loss_per_oscillation}
            \Delta E_{\text{i}}=\int_0^\tau  L_{\text{i}}\eta_{\text{BS}}\dot{s}_{\text{i}} \frac{\mathrm{d} s_{\text{i}}}{\mathrm{d}t}\mathrm{d}t=\frac{1}{2}\tau L_{\text{i}} \frac{\Phi_0^2}{2 \pi \xi^2 \rho_n}\left(\frac{k_{\text{m,i}}}{k_{\text{v,i}}}\right)^2a^2\omega^2_{\mathrm{m}},
        \end{align}
        where $L_{\text{i}}$ is the length of \textit{i}-th vortex.

        Collectively, the motion of the magnet leads to the motion of all vortices that support mechanical modes. Figure~\ref{Appen_fig_6} shows a schematic of such a coupled system. The displacement of the magnet (blue circle) $x$ drags \textit{i}-th vortex (red box) from its equilibrium point in its pinning center by $s_{\text{i}}$. The equation of the motion is
        \begin{align}     m\ddot{x}+m(\gamma_{\mathrm{M}}+\Gamma_{\mathrm{V}})\dot{x}+\sum_i k_{\mathrm{m,i}} (x-s_i)=F_{\mathrm{th}},
            \end{align}
        where $m$ is the mass of the magnet, $\gamma_{\mathrm{M}}$ is the intrinsic mechanical dissipation rate, $\Gamma_{\text{V}}$ is the additional dissipation caused by vortex motion, $k_{\text{m,i}}$ is the effective spring constant of the magnet from the \textit{i}-th vortex, $s_i$ is the displacement of \textit{i}-th vortex, and $F_{\text{th}}$ is the thermal noise. Displacements of each vortex are in phase with the far-off resonant drive from the mechanical motion as $s_i={k_{\text{m,i}}}/({k_{\text{v,i}}+k_{\text{m,i}}})x$, where $k_{\mathrm{v,i}}$ is the effective spring constant for the \textit{i}-th vortex in the superconductor.
        Therefore, the effective spring constant is $k\approx \sum_i k_{\mathrm{m}, \mathrm{i}}-\sum_i {k_{\mathrm{m}, \mathrm{i}}^2}/{k_{\mathrm{v,i}}}$,
        where the approximation is taken by assuming $k_{\text{v,i}}\gg k_{\text{m,i}}$ and the corresponding mechanical frequency is
        \begin{align}
            \omega'_{\mathrm{m}}\approx\omega_{\mathrm{m}}(1-\frac{1}{2}\sum_i \frac{k_{\mathrm{m}, \mathrm{i}}^2}{k k_{\mathrm{v,i}}}),
        \end{align}
        where $\omega_{\mathrm{m}}=\sqrt{\sum_ik_{\mathrm{m,i}}/m}$ is the intrinsic mechanical frequency with fixed vortices.

        The dissipation is induced by excited currents in the normal core of the vortex as described in the Bardeen-Stephen model~\cite{bardeen_1965_theory_PR}, which can be phenomenologically associated with a viscous force $f_{\text{BS}}=-L\eta_{\text{BS}}\dot{s}$ on the moving vortex, where $\eta_{\text{BS}}={\Phi_0^2}/({2 \pi \xi^2 \rho_n})$ is the Barden-Stephen viscosity with $\rho_n$ being the resistivity of the electron fluid in the vortex core, often taken to be the normal state resistivity, $L$ is the length of the vortex. Therefore, the effective magnet damping rate by vortex motion $\Gamma_{\mathrm{V}}$ is
        \begin{align}         \Gamma_{\text{V}}\approx\sum_i L_i\eta_{\text{BS}}\left({k_{\text{m,i}}}/{k_{\text{v,i}}}\right)^2/m.
        \end{align}           

    \subsection{Temperature dependence}
        \label{sec:Temperature dependence}  
        The YBCO film we used in the experiment is around \SI{270}{\nano\meter} thick, which is larger than the penetration depth $\lambda$ of YBCO at \SI{6}{\kelvin}. 
        A more detailed consideration should include the effect of penetration depth. The surface current is distributed along the $z$ direction as
        \begin{align}
            j(z)=\frac{\Delta J}{\lambda}e^{-z/\lambda}, 
        \end{align}
        where $\lambda$ is the in-plane penetration depth. Similarly, the Lorentz force per unit length is also distributed as $f_{\text{L}}(z)={F_{\mathrm{V}}}/{\lambda}e^{-z/\lambda}$ with $F_{\mathrm{V}}$ being the force on a single vortex. We can define the elasticity of the collective effects from pinning centers as
        \begin{align}
            c_{\text{v}}=\tilde{k}_{\text{v}}\lambda_0,
        \end{align}
        where $\tilde{k}_{\text{v}}$ is the spring constant of a single pinning center, and $\lambda_0$ is the mean distance between pinning centers. Therefore, the effective spring constant is
        \begin{align}
            \frac{1}{k_{\text{v,i}}}=\int_{0}^{\infty}\frac{1}{c_{\text{v}}}\times e^{-z/\lambda} \mathrm{d}z=\frac{1}{\tilde{k}_{\text{v}}}\frac{\lambda}{\lambda_0}.
        \end{align}
        This can be understood as several pinning centers connected by a single vortex in series, where ${\lambda}/{\lambda_0}$ is the average number of pinning centers per vortex.\\
        The displacement $\Delta s_\text{i}$ of the vortex follows
        \begin{align}
            \Delta s_\text{i}(z)=\int_{z}^{\infty}\frac{F_{\mathrm{V}}\lambda_0}{\lambda c}e^{-z/\lambda}\mathrm{d}z=\frac{F_{\mathrm{V}}\lambda_0}{c}e^{-z/\lambda}.
        \end{align}
        Therefore, the effective vortex-induced damping is
        % \begin{widetext}
        \begin{align}
            \begin{split}\label{eq:dissipation_expression}
                \Gamma_{\text{V}}&\approx
                \sum_{i}\frac{1}{2}\lambda\frac{\Phi_0^2}{2 \pi \xi^2 \rho_n}\left( \frac{k_{\text{m,i}}}{\tilde{k}_{\text{v}}+k_{\text{m,i}}}\right)^2,
            \end{split}
        \end{align}
        and the frequency of the mechanical mode is
        \begin{align}\label{eq:freq_expression}
            \omega'_{\text{m}}=\sqrt{\frac{\sum_ik_{\text{m,i}}-\sum_i {k_{\mathrm{m}, \mathrm{i}}^2}/{k_{\mathrm{v}, \mathrm{i}}}}{m}},
        \end{align}
        We can further simplify  Eq.~\eqref{eq:dissipation_expression} and Eq.~\eqref{eq:freq_expression} by assuming all vortices are identical. So we have
        \begin{align}\label{eq:freq_temp_dependence}
            \omega'_{\text{m}}(T)\approx\omega_{\text{m}}(1-\frac{1}{2}\frac{\bar{k}_{\text{m}}}{\bar{k}_{\text{v}}(T)}\frac{\lambda(T)}{\lambda_0}),
        \end{align} 
        and 
        \begin{align}\label{eq:dissipation_expression_avg}
            \Gamma_{\mathrm{V}}(T)\approx N\frac{1}{2}\frac{\lambda(T)}{m}\frac{\Phi_0^2}{2 \pi \xi(T)^2 \rho_n}\left( \frac{\bar{k}_{\text{m}}}{\bar{k}_{\text{v}}(T)}\right)^2,
        \end{align}
        where $\bar{k}_{\text{m}}$ is the average spring constant for the magnet per vortex, $\bar{k}_{\text{v}}$ is the average spring constant of pinning centers, $\omega_{\text{m}}=\sqrt{\sum k_{\text{m,i}}/m}$.\\ 

        Hence, the temperature dependence of mechanical response are related to the temperature dependence of three superconductor parameters:
        \begin{description}
            \item[London Penetration Depth $\lambda$]{
            YBCO is a d-wave superconductor, meaning its energy gap has nodes (zeroes) on the Fermi surface. Because of these nodes, quasiparticles are thermally excited even at very low temperatures, leading to a linear increase in $\lambda(T)$ as~\cite{Hardy_1993_Precision_PRL}
            \begin{align}\label{eq:linear_relation}
                \lambda(T)= \lambda(0)(1+\beta T/T_{\mathrm{c}}),
            \end{align}
            where $\beta$ is a constant. At higher-temperature, near $T_{\mathrm{c}}$, the following power law applies
            \begin{align}
                \lambda(T)= \lambda(0)(1- T/T_{\mathrm{c}})^{-1/2}.
            \end{align}
            }
            In our case, we have measured mechanical frequencies and dissipations up to 80~K, at which the penetration depth should be greater than the sample thickness (270 nm), so the penetration depth is approximated as a constant at higher temperatures.
            
            \item[Coherence Length $\xi$]{
            We take the Ginzburg–Landau model to account for the coherence length
            \begin{align}
                \xi(T)=\xi_0(1-T/T_{\text{c}})^{-1/2},
            \end{align}
            where $T_{\mathrm{c}}$ is the critical temperature, and $\xi_0$ is the coherence length at 0 K.
            }
            
            \item[Labusch (pinning) parameter $\alpha$]{Using the basic Ginzburg-Landau scalings near $T_{\mathrm{c}}$, we have the Labusch parameter $\alpha$ (\textit{i.e.}, the spring constant of pinning potential) 
            \begin{align}
                \alpha(T)\propto\alpha_0(1-T/T_{\mathrm{c}})^{p},
            \end{align}
            }
            where $p$ is taken as 1. 
        \end{description}    
        Therefore, we have
        \begin{align}
            \omega'_{\text{m}}&=\omega_{\text{m}}(1-b\frac{1}{1-t}),
            \label{eq:freq_vs_temp_fit}\\
            \Gamma_{\text{V}}&=\Gamma_0\frac{1+\beta t}{1-t}\label{eq:linewidth_vs_temp_fit},
        \end{align}
        where $t=T/T_{\mathrm{c}}$ is normalized temperature, $b$ and $\beta$ are dimensionless fitting parameters, and $\Gamma_0$ is the dissipation at 0~K. Specifically, we treat $\lambda$ as a constant for the mode frequency, because the measured frequency changes are more significant at high temperatures ($>$~40~K). However, for dissipation rates, the most accurate data are measured at low temperatures ($<$ 60~K), so we use Eq.~\eqref{eq:linear_relation} for the penetration depth $\lambda(T)$.
           
        The comparison between the model and experimental results is shown in Fig.~3(a,b) in the main text. The fitting to the measured linewidth suggests the dissipation rate is almost proportional to the YBCO temperature at low temperatures, which is not consistent with the results in Ref.~\cite{Hardy_1993_Precision_PRL}. Further investigation is still needed to explain this result.
     
    \subsection{Measurements of temperature dependence at different \textit{B}}
        \label{sec:Temperature dependence at different B}
        We measure the temperature dependence of frequency and damping rate at various applied external magnetic fields.
        %%%%%%%%%%%%%%%%%%%%%%%%%%%%%%%%%%%%%%%%%%%%%%%%%%%%%%%%%%%%%%%%%%
        \begin{figure}
            \centering
            \includegraphics[width=1\linewidth]{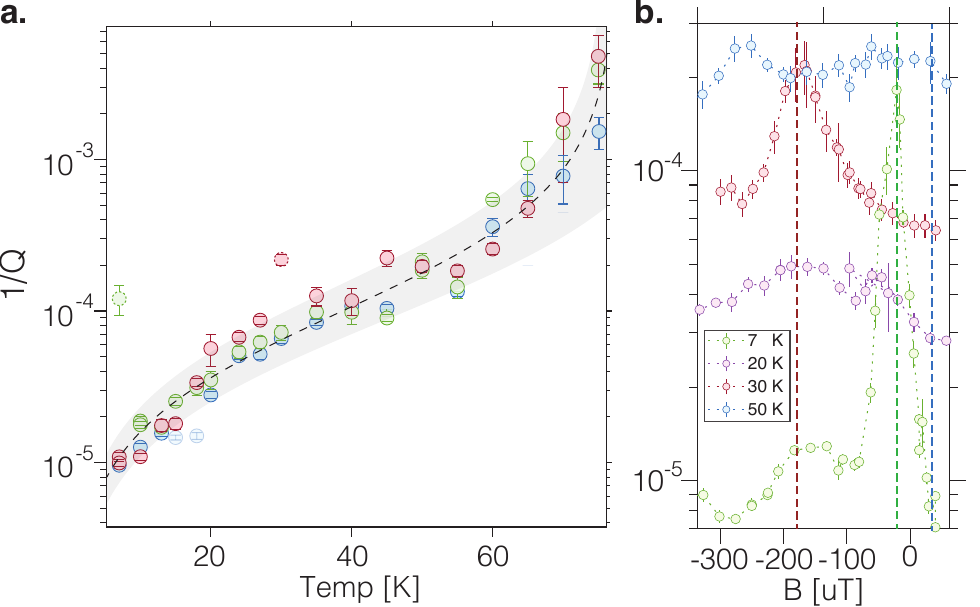}
            \caption{\textbf{Temperature dependence of mechanical dissipation rates. }
            (a) Measured inverse mechanical quality factor $1/Q$ as a function of temperature. Error bars represent 2 s.d. determined from fits. Colors of circles correspond to sweeps at different $B$ fields that are highlighted by the dashed line of the same color, as shown in (b). The black dashed line is the global fit to Eq.~\eqref{eq:linewidth_vs_temp_fit}, excluding a few outliers highlighted by dashed circles. The gray shaded region represents the 95\% confidence interval of the fit. 
            (b) Inverse mechanical quality factors as a function of $B$ at 7, 20, 30, and 50~K. Each data point is extracted from fits to the average of 20 ringdowns. Error bars represent 2 s.d. determined from fits.}
            \label{Appen_fig_7}
        \end{figure}
        %%%%%%%%%%%%%%%%%%%%%%%%%%%%%%%%%%%%%%%%%%%%%%%%%%%%%%%%%%%%%%%%%%

        \begin{description}
            \item[Frequency]{
            The temperature dependence of mechanical resonance frequencies is measured at 3 different $B$ fields (-87, 65, 104 \SI{}{\micro\tesla}) and on a second levitation (detailed in Appendix~\ref{sec: Repeatability of an-isotropic damping rate}) of the same magnet at the same location. 
           
            The measured frequency is fit to Eq.~\eqref{eq:freq_vs_temp_fit} to extract parameter $b$, shown in Fig.~\ref{fig:5}. A detailed discussion is given in Sec.~\ref{sec:Mechanical-vortex coupling} of the main text.  

            According to Eq.~\eqref{eq:freq_temp_dependence}, the value of $b$ represents  $\bar{k}_{\text{m}}/2\bar{k}_{\text{v}}(0)$, where $\bar{k}_{\text{m}}$ can be estimated by $m\omega^2_{\mathrm{m}}/N$ with $N$ being the number of vortices. So we have 
            \begin{align}
                \bar{k}_{\text{v}}(0)=\frac{2m\pi^2}{Nb/f_0^2}.
            \end{align}
            The mass of the particle is around \SI{100}{\pico\gram}, and the effective number of vortices is around 4 (see Appendix~\ref{sec:Estimation_of_number_vortices}), and $b/f_0^2\approx3.6\times10^{-11}~\mathrm{Hz}^{-2}$ for $x$ mode. So we have $\bar{k}_{\text{v}}(0)\approx0.014~\mathrm{N/m}$. Thus, the mean Labusch parameter is $\bar{k}_{\text{v}}(0)/L=~$\SI{51}{\kilo\newton\per\meter^2} with $L=270~\mathrm{nm}$ given by the thickness of the YBCO, reasonably consistent with literature values~\cite{doyle_1993_direct_PRL, pesetski_2000_experimental_prb}. 
            }
            
            \item[Dissipation]{The dissipation rates depend on the temperature and applied magnetic field $B$. We repeat more than 20 ringdown measurements at each $B$ when the YBCO is thermalized at 7, 20, 30, and 50~K. The measured mean inverse of the mechanical quality factor $1/Q$ is presented in Fig.~\ref{Appen_fig_7}(b). We find that overall, the dissipation rate increases with increasing temperature as we expect. It also exhibits a strong magnetic field dependence around certain $B$ fields at each temperature. At \SI{7}{\kelvin}, we observe a more than 40-fold increase in mechanical dissipation. A similar peak feature appears at \SI{30}{\kelvin}, but at a different magnetic field $B$ and with only a four-fold increase in mechanical dissipation, yielding a peak dissipation rate comparable to that observed at \SI{7}{\kelvin}.
        
            To elucidate the complex temperature dependence of dissipation, we further repeat ringdown measurements with increasing temperature at three representative $B$ fields. Figure~\ref{Appen_fig_7}(a) shows the measured average $1/Q$. Each point is color-coded to match the corresponding dashed line in Fig.~\ref{Appen_fig_7}(b). The overall temperature dependence of the three sweeps is consistent with each other, consistent with the model in Eq.~\eqref{eq:linewidth_vs_temp_fit}. Peak features in this $B$ sweep manifest into a few outliers, highlighted by dashed circles. 
            
            An applied magnetic field exerts forces on individual vortices, while increasing temperature softens the pinning potential. We interpret the complex magnetic field and temperature dependence of the mechanical dissipation as arising from the interplay between the temperature-dependent vortex restoring force and the external magnetic force, which together lead to vortex reconfiguration. We attribute the sudden rise in dissipation observed in the $B$ sweep to a complete depinning process of a single vortex, which also accounts for similar maximal dissipation rates in $B$ sweep at \SI{7}{\kelvin} and \SI{30}{\kelvin}.
            }
        \end{description}

    %%%%%%%%%%%%%%%%%%%%%%%%%%%%%%%%%%%%%%%%%%%%%%%%%%%%%%%%%%%%%%%%%
    \begin{figure}
        \centering
        \includegraphics[width=1\linewidth]{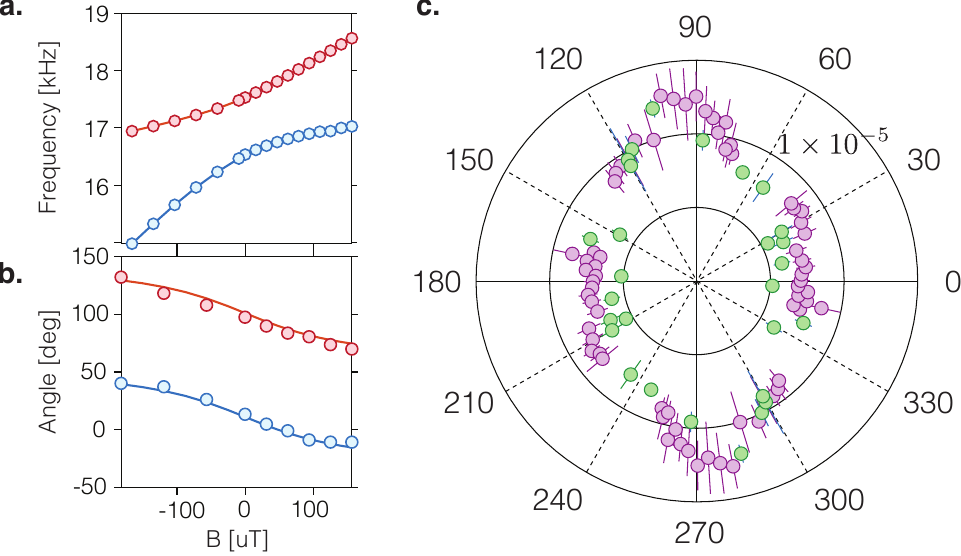}
        \caption{\textbf{Angle-dependent mechanical dissipation of a separate levitation. }
        (a,b) Frequencies and motion angles of two lateral modes as a function of $B$. Solid lines
        are fits to a linear coupled mode theory described in the main text. 
        (c) Inverse mechanical quality factor $1/Q$ as a function of motional angles. Purple circles are results of the second levitation, as described in Appendix~\ref{sec: Repeatability of an-isotropic damping rate}. For comparison, green circles are the results shown in the main text. Error bars are 2 s.d. determined from fits and statistical uncertainties.}
        \label{Appen_fig_8}
    \end{figure}
    %%%%%%%%%%%%%%%%%%%%%%%%%%%%%%%%%%%%%%%%%%%%%%%%%%%%%%%%%%%%%%%%% 
    
    \subsection{Repeatability of anisotropic mechanical dissipation}
        \label{sec: Repeatability of an-isotropic damping rate}  
        For thoroughness, we repeat the angle-dependent dissipation measurements after reconfiguring the magnet trap. This was done by warming the cryostat up to \SI{80}{\kelvin}, then applying an external magnetic field, and eventually cooling it back to \SI{6}{\kelvin} and removing the applied magnetic field. Such a procedure sufficiently scrambles the vortex configuration, leading to different zero-field translational mode frequencies and directions at \SI{6}{\kelvin}. The $x$ and $y$ mode frequencies shift from \SI{14.091}{\kilo\hertz} and \SI{12.841}{\kilo\hertz} to \SI{17.530}{\kilo\hertz} and \SI{16.634}{\kilo\hertz}, respectively. We re-approach the avoided crossing between two in-plane translational modes as shown in Fig.~\ref{Appen_fig_8}(a,b), now centered around \SI{30.4}{\micro\tesla} compared to \SI{116.2}{\micro\tesla} as shown in Fig.~3 of the main text. The effective linear coupling rate increases from $g_0/2\pi=~84.2\pm1.4~\text{Hz}$ to $g_0/2\pi=~442\pm13~\text{Hz}$. The new angular dependence of mechanical dissipation is shown as purple circles in Fig.~\ref{Appen_fig_8}(c), aligned with the results (green circles) in the main text. Given the reconfiguration of the trapping potential, evidenced by changes in motional frequency, magnet orientation, and mode coupling, the reproducibility rules out that this angle-dependent dissipation is a result of the $B$ field dependence discussed in Appendix~\ref{sec:Temperature dependence at different B} or the eddy-current-induced dissipation discussed in Appendix~\ref{sec:eddy_current_estimate}.

    \subsection{Strong pinning model}
        \label{sec:Strong pinning theory}
        A lateral displacement $x$ of the vortex segment at the shallow pinning layer increases the vortex length in the superconductor (see Fig.~\ref{fig:4}(b)) by
        $L=L_0\left(1+\left({x}/{L_1}\right)^2/2\right)
        $. Combining the first term and the third term in Eq.~\eqref{eq:Gibbs_energy_pinning} yields an effectively shifts $r_0$ by $\Delta r= {\Phi_0BL_1^2}/{\mu_0\epsilon_{\text{L}}L_0}$.
        When the Gibbs energy portrays a double-well, the vortex requires a threshold energy to overcome the depinning barrier, transitioning from local oscillations to complete pinning-depinning processes~\cite{buchacek_2019_strong_PRB}. For a small pinning size ($r_0\gg r_{\text{us}}\gg \epsilon$,~where $r_{\text{us}}$ is the location of the metastable point), the threshold energy can be approximated as  
        \begin{align}\label{eq:threshold_energy}
            E'(B)\approx (aB+b)^{2/3}+c,
        \end{align}
        where $a,b,c$ are parameters determined from Eq.~\eqref{eq:Gibbs_energy_pinning}. Equation~\eqref{eq:threshold_energy} is used to fit data in Fig.~4(d). \\
   
\section{LIMITS OF MAGNETIC LEVITATION SYSTEMS}
    \label{sec:Vortex motion limited magnetic levitation system}
    In this section, we hypothesize a limit to the mechanical quality factor of the levitated magnet above $\operatorname{type-II}$ superconductors. Such a limit can also be extended to other magnetic levitated systems with or without superconductors. 
    
    Consider a magnet of radius $r$ with remanence $B_{\text{r}}$ levitating above a $\operatorname{type-II}$ superconductor of height $z$. A single vortex provides an effective spring constant $k_{\text{m}}$ given by
    \begin{align}
        k_{\text{m}}\approx\frac{3\Phi_0VB_{\text{r}}}{\pi z^4},
    \end{align}
    where $V$ is the volume of the magnet.
    The number of vortices should follow
    \begin{align}
        N\propto B_r(\frac{r}{z})^3z^2\propto B_r\frac{r^3}{z},
    \end{align}
    where $({r}/{z})^3$ evaluates the magnetic field strength on the superconductor surface and $z^2$ accounts for the surface area of the trap source.
    Therefore, the mechanical frequency follows
    \begin{align}
        \omega_{\mathrm{m}}\approx\sqrt{\frac{Nk_{\text{m}}}{m}}\propto\frac{B_{\text{r}}}{\sqrt{\rho}}\frac{1}{r}\left(\frac{r}{z}\right)^{5/2},
    \end{align}
    where $\rho$ is the density of the magnet. This result agrees with the conclusion in other Refs~\cite{gieseler_2020_single_PRL,kordyuk_1998_magnetic_JAP}. The dissipation rates of mechanical motion follow
    \begin{align}
        \gamma_{m}\propto \frac{N}{\rho V}{k_{\text{m}}^2}\propto \frac{r^6}{z^9}\frac{B_\text{r}^3}{\rho}.
    \end{align}
    Therefore, the mechanical quality factor follows
    \begin{align}
        Q=\frac{\omega_{\text{m}}}{\gamma_{\mathrm{M}}}\propto \frac{\sqrt{\rho}}{B^2_{\text{r}}}\frac{z^{6.5}}{r^{4.5}}\propto\frac{1}{B_{\text{r}}}\frac{1}{\omega_{\mathrm{m}}}\left(\frac{r}{z}\right)^{-4}r.
    \end{align}
    Hence, for given superconductor properties and a relatively similar geometric ratio $r/z$, we can define a parameter 
    \begin{align}
        \alpha=\frac{fQ}{r}\leq \alpha_0,
    \end{align}
    where $\alpha_0$ is mainly determined by the properties of pinning sites and the magnet remanence. 
    This relation indicates a tradeoff between the mechanical frequency and its corresponding quality factor. 
    
    Such a constraint may extend to other magnetic levitation systems, such as type-I superconductors levitated in a magnetic trap~\cite{hofer_2023_high_PRL,gutierrez_2023_superconducting_PRApplied} or diamagnetically levitated particles~\cite{brown_2023_superfluid_PRL,Tian_2024_feedback_APL}. Even though these systems do not explicitly involve vortices, the scaling discussed above generally captures how a magnetic field constrains the levitated system and how the particle’s motion produces backaction on the field source. In our case, this backaction drives vortex motion, whereas in other systems it may simply generate additional oscillating currents leading to ohmic loss. For example, considering a diamagnetically levitated system, the resonance frequency is independent of the size of the particle. It relies on the system dimension $z$ as $f\propto {1}/{z}$. Thus, the net force $F$ on the system scales as
    \begin{align}
        F\propto mf^2\propto r^3f^2.
    \end{align}
    The net force is generated by the source of the magnetic field, with an induced field density from the particle given by 
    \begin{align}
        E\propto \frac{F}{z^2}(\frac{z}{r})^3,
    \end{align}
    where $z^2$ accounts for the surface area of the system and $({z}/{r})^3$ accounts for the scaling of effective dipole fields from the levitated particle. Therefore, the dissipation rate scales as
    \begin{align}
        \gamma_{\mathrm{M}}=\frac{\Gamma}{m}\propto \frac{E^2z^2}{r^3},
    \end{align}
    where $E^2$ term represents ohmic-like loss, $z^2$ represents the total area. Therefore, such a system also exhibits a similar scaling:
    \begin{align}
        \frac{fQ}{r}\propto(\frac{r}{z})^2.
    \end{align}

\bibliographystyle{apsrev4-2}
\bibliography{main}

@article{gonzalez_2021_levitodynamics_science,
  title={Levitodynamics: Levitation and control of microscopic objects in vacuum},
  author={Gonzalez-Ballestero, Carlos and Aspelmeyer, Markus and Novotny, Lukas and Quidant, Romain and Romero-Isart, Oriol},
  journal={Science},
  volume={374},
  number={6564},
  pages={eabg3027},
  year={2021},
  publisher={American Association for the Advancement of Science}
}

@article{delic_2020_cooling_nature,
  title={Cooling of a levitated nanoparticle to the motional quantum ground state},
  author={Deli{\'c}, Uro{\v{s}} and Reisenbauer, Manuel and Dare, Kahan and Grass, David and Vuleti{\'c}, Vladan and Kiesel, Nikolai and Aspelmeyer, Markus},
  journal={Science},
  volume={367},
  number={6480},
  pages={892--895},
  year={2020},
  publisher={American Association for the Advancement of Science}
}

@article{magrini_2021_real_science,
  title={Real-time optimal quantum control of mechanical motion at room temperature},
  author={Magrini, Lorenzo and Rosenzweig, Philipp and Bach, Constanze and Deutschmann-Olek, Andreas and Hofer, Sebastian G and Hong, Sungkun and Kiesel, Nikolai and Kugi, Andreas and Aspelmeyer, Markus},
  journal={Nature},
  volume={595},
  number={7867},
  pages={373--377},
  year={2021},
  publisher={Nature Publishing Group UK London}
}

@article{Rossi_2025_quantum_prl,
  title = {Quantum Delocalization of a Levitated Nanoparticle},
  author = {Rossi, M. and Militaru, A. and Carlon Zambon, N. and Riera-Campeny, A. and Romero-Isart, O. and Frimmer, M. and Novotny, L.},
  journal = {Phys. Rev. Lett.},
  volume = {135},
  issue = {8},
  pages = {083601},
  numpages = {7},
  year = {2025},
  month = {Aug},
  publisher = {American Physical Society},
  doi = {10.1103/2yzc-fsm3},
}

@article{kamba_2025_quantum_science,
  title={Quantum squeezing of a levitated nanomechanical oscillator},
  author={Kamba, Mitsuyoshi and Hara, Naoki and Aikawa, Kiyotaka},
  journal={Science},
  volume={389},
  number={6766},
  pages={1225--1228},
  year={2025},
  publisher={American Association for the Advancement of Science}
}

@article{lewandowski_2021_high_PRApplied,
  title={High-sensitivity accelerometry with a feedback-cooled magnetically levitated microsphere},
  author={Lewandowski, Charles W and Knowles, Tyler D and Etienne, Zachariah B and D’Urso, Brian},
  journal={Phys. Rev. Appl.},
  volume={15},
  number={1},
  pages={014050},
  year={2021},
  publisher={APS}
}

@article{Tian_2024_feedback_APL,
  title={Feedback cooling of an insulating high-$\mathrm{Q}$ diamagnetically levitated plate},
  author={Tian, S and Jadeja, K and Kim, D and Hodges, A and Hermosa, GC and Cusicanqui, C and Lecamwasam, R and Downes, JE and Twamley, J},
  journal={Appl. Phys. Lett.},
  volume={124},
  number={12},
  year={2024},
  publisher={AIP Publishing}
}

@article{bose_2017_spin_PRL,
  title={Spin entanglement witness for quantum gravity},
  author={Bose, Sougato and Mazumdar, Anupam and Morley, Gavin W and Ulbricht, Hendrik and Toro{\v{s}}, Marko and Paternostro, Mauro and Geraci, Andrew A and Barker, Peter F and Kim, MS and Milburn, Gerard},
  journal={Phys. Rev. Lett.},
  volume={119},
  number={24},
  pages={240401},
  year={2017},
  publisher={APS}
}

@article{gieseler_2020_single_PRL,
  title={Single-spin magnetomechanics with levitated micromagnets},
  author={Gieseler, Jan and Kabcenell, Aaron and Rosenfeld, Emma and Schaefer, JD and Safira, Arthur and Schuetz, Martin JA and Gonzalez-Ballestero, Carlos and Rusconi, Cosimo C and Romero-Isart, Oriol and Lukin, Mikhail D},
  journal={Phys. Rev. Lett.},
  volume={124},
  number={16},
  pages={163604},
  year={2020},
  publisher={APS}
}

@article{hofer_2023_high_PRL,
  title={High-$\mathrm{Q}$ magnetic levitation and control of superconducting microspheres at millikelvin temperatures},
  author={Hofer, Joachim and Gross, Rudolf and Higgins, Gerard and Huebl, Hans and Kieler, Oliver F and Kleiner, Reinhold and Koelle, Dieter and Schmidt, Philip and Slater, Joshua A and Trupke, Michael and others},
  journal={Phys. Rev. Lett.},
  volume={131},
  number={4},
  pages={043603},
  year={2023},
  publisher={APS}
}

@article{gutierrez_2023_superconducting_PRApplied,
  title={Superconducting microsphere magnetically levitated in an anharmonic potential with integrated magnetic readout},
  author={Gutierrez Latorre, Mart{\'\i} and Higgins, Gerard and Paradkar, Achintya and Bauch, Thilo and Wieczorek, Witlef},
  journal={Phys. Rev. Appl.},
  volume={19},
  number={5},
  pages={054047},
  year={2023},
  publisher={APS}
}

@article{smit_2025_superconducting_arXiv,
  title={Superconducting flux concentrator coils for levitation of particles in the Meissner state},
  author={Smit, Robert and Janse, Martijn and van der Bent, Eli and de Jong, Thijmen and Heeck, Kier and Plugge, Jaimy and Oosterkamp, Tjerk and Hensen, Bas},
  journal={PNAS nexus},
  volume={5},
  number={3},
  pages={pgag072},
  year={2026},
  publisher={Oxford University Press US}
}

@article{chen_2007_two_natphy,
  title={Two-dimensional vortices in superconductors},
  author={Chen, Bo and Halperin, WP and Guptasarma, Prasenjit and Hinks, DG and Mitrovi{\'c}, VF and Reyes, AP and Kuhns, PL},
  journal={Nat. Phys.},
  volume={3},
  number={4},
  pages={239--242},
  year={2007},
  publisher={Nature Publishing Group UK London}
}

@article{auslaender_2009_mechanics_natphy,
  title={Mechanics of individual isolated vortices in a cuprate superconductor},
  author={Auslaender, Ophir M and Luan, Lan and Straver, Eric WJ and Hoffman, Jennifer E and Koshnick, Nicholas C and Zeldov, Eli and Bonn, Douglas A and Liang, Ruixing and Hardy, Walter N and Moler, Kathryn A},
  journal={Nat. Phys.},
  volume={5},
  number={1},
  pages={35--39},
  year={2009},
  publisher={Nature Publishing Group UK London}
}

@article{eley_2017_universal_natmat,
  title={Universal lower limit on vortex creep in superconductors},
  author={Eley, Serena and Miura, Masashi and Maiorov, Boris and Civale, Leonardo},
  journal={Nat. Mat.},
  volume={16},
  number={4},
  pages={409--413},
  year={2017},
  publisher={Nature Publishing Group UK London}
}

@article{buchacek_2019_experimental_PRB,
  title={Experimental test of strong pinning and creep in current-voltage characteristics of type-$\mathrm{II}$ superconductors},
  author={Buchacek, Martin and Xiao, Zhi-Li and Dutta, Surajit and Andrei, Eva Y and Raychaudhuri, Pratap and Geshkenbein, Vadim B and Blatter, Gianni},
  journal={Phys. Rev. B},
  volume={100},
  number={22},
  pages={224502},
  year={2019},
  publisher={APS}
}

@article{maksymowych_2025_spectral_PRApplied,
  title={Spectral diffusion of nanomechanical resonators due to single quantum defects},
  author={Maksymowych, MP and Yuksel, M and Hitchcock, OA and Lee, NR and Mayor, FM and Jiang, W and Roukes, ML and Safavi-Naeini, AH},
  journal={Phys. Rev. Appl.},
  volume={24},
  number={4},
  pages={044066},
  year={2025},
  publisher={APS}
}

@article{zhang_2024_acceptor_PRL,
  title={Acceptor-induced bulk dielectric loss in superconducting circuits on silicon},
  author={Zhang, Zi-Huai and Godeneli, Kadircan and He, Justin and Odeh, Mutasem and Zhou, Haoxin and Meesala, Srujan and Sipahigil, Alp},
  journal={Phys. Rev. X},
  volume={14},
  number={4},
  pages={041022},
  year={2024},
  publisher={APS}
}

@article{faoro_2015_interacting_PRB,
  title={Interacting tunneling model for two-level systems in amorphous materials and its predictions for their dephasing and noise in superconducting microresonators},
  author={Faoro, Lara and Ioffe, Lev B},
  journal={Phys. Rev. B},
  volume={91},
  number={1},
  pages={014201},
  year={2015},
  publisher={APS}
}

@article{kennedy_2025_josephson_PRL,
  title={Josephson Junction Tuning Described by Depinning Physics},
  author={Kennedy, Oscar W and Cole, Jared H and Shelly, Connor D},
  journal={Phys. Rev. Lett.},
  volume={135},
  number={19},
  pages={196202},
  year={2025},
  publisher={APS}
}

@article{nambisan_2025_quantum_arXiv,
  title={Quantum coherent manipulation and readout of superconducting vortex states},
  author={Nambisan, Ameya and G{\"u}nzler, Simon and Rieger, Dennis and Gosling, Nicolas and Geisert, Simon and Carpentier, Victor and Zapata, Nicolas and Field, Mitchell and Milo{\v{s}}evi{\'c}, Milorad V and Lopez, Carlos A Diaz and others},
  journal={Nature},
  volume={653},
  number={8113},
  pages={63--67},
  year={2026},
  publisher={Nature Publishing Group UK London}
}

@article{buchacek_2019_strong_PRB,
  title={Strong pinning theory of thermal vortex creep in type-$\mathrm{II}$ superconductors},
  author={Buchacek, Martin and Willa, Roland and Geshkenbein, Vadim B and Blatter, Gianni},
  journal={Phys. Rev. B},
  volume={100},
  number={1},
  pages={014501},
  year={2019},
  publisher={APS}
}

@article{cleland_2024_studying_NatCom,
  title={Studying phonon coherence with a quantum sensor},
  author={Cleland, Agnetta Y and Wollack, E Alex and Safavi-Naeini, Amir H},
  journal={Nat. Commun.},
  volume={15},
  number={1},
  pages={4979},
  year={2024},
  publisher={Nature Publishing Group UK London}
}

@article{guttinger_2017_energy_NatNan,
  title={Energy-dependent path of dissipation in nanomechanical resonators},
  author={G{\"u}ttinger, Johannes and Noury, Adrien and Weber, Peter and Eriksson, Axel Martin and Lagoin, Camille and Moser, Joel and Eichler, Christopher and Wallraff, Andreas and Isacsson, Andreas and Bachtold, Adrian},
  journal={Nat. Nano.},
  volume={12},
  number={7},
  pages={631--636},
  year={2017},
  publisher={Nature Publishing Group UK London}
}

@article{kordyuk_1998_magnetic_JAP,
  title={Magnetic levitation for hard superconductors},
  author={Kordyuk, Alexander A},
  journal={Jour. of Appl. Phys.},
  volume={83},
  number={1},
  pages={610--612},
  year={1998},
  publisher={American Institute of Physics}
}

@article{hansen_2025_optical_arXiv,
  title={Optical interferometric readout of a magnetically levitated superconducting microsphere},
  author={Hansen, Jannek J and Minniberger, Stefan and Ilk, Dominik and Asenbaum, Peter and Higgins, Gerard and Povey, Rhys G and Schmidt, Philip and Hofer, Joachim and Claessen, R{\'e}mi and Aspelmeyer, Markus and others},
  journal={Phys. Rev. Appl.},
  volume={25},
  number={4},
  pages={044080},
  year={2026},
  publisher={APS}
}

@article{labusch_1969_elastic_PSS,
  title={Elastic constants of the fluxoid lattice near the upper critical field},
  author={Labusch, R},
  journal={Phys. Sta. Sol. (b)},
  volume={32},
  number={1},
  pages={439--442},
  year={1969},
  publisher={Wiley Online Library}
}

@article{campbell_1969_response_JPC,
  title={The response of pinned flux vortices to low-frequency fields},
  author={Campbell, AM},
  journal={Jour. of Phys. C: Sol. St. Phys.},
  volume={2},
  number={8},
  pages={1492},
  year={1969},
  publisher={IOP Publishing}
}

@article{doyle_1993_direct_PRL,
  title={Direct observation of intrinsic pinning in $\mathrm{YBCO}$ thin films},
  author={Doyle, RA and Campbell, AM and Somekh, RE},
  journal={Phys. Rev. Lett.},
  volume={71},
  number={25},
  pages={4241},
  year={1993},
  publisher={APS}
}

@article{compton_2006_dynamics_PRL,
  title={Dynamics of a pinned magnetic vortex},
  author={Compton, RL and Crowell, PA},
  journal={Phys. Rev. Lett.},
  volume={97},
  number={13},
  pages={137202},
  year={2006},
  publisher={APS}
}

@article{mehrnia_2024_observation_JMMM,
  title={Observation of pinning and pinning evasion dynamics of a magnetic vortex core},
  author={Mehrnia, Mahdi and Berezovsky, Jesse},
  journal={Jour. of Mag. and Mag. Mat.},
  volume={593},
  pages={171885},
  year={2024},
  publisher={Elsevier}
}

@article{bardeen_1965_theory_PR,
  title={Theory of the motion of vortices in superconductors},
  author={Bardeen, John and Stephen, MJ},
  journal={Phys. Rev.},
  volume={140},
  number={4A},
  pages={A1197},
  year={1965},
  publisher={APS}
}

@article{palau_2006_crossover_PRL,
  title={Crossover between channeling and pinning at twin boundaries in $\mathrm{YBa_2Cu_3O_7}$ thin films},
  author={Palau, A and Durrell, JH and MacManus-Driscoll, JL and Harrington, S and Puig, T and Sandiumenge, F and Obradors, <? format?> X and Blamire, MG},
  journal={Phys. Rev. Lett.},
  volume={97},
  number={25},
  pages={257002},
  year={2006},
  publisher={APS}
}

@article{pereg_2004_absolute_PRB,
  title={Absolute values of the London penetration depth in $\mathrm{YBa2Cu3O_{6+ y}}$ measured by zero field ESR spectroscopy on Gd doped single crystals},
  author={Pereg-Barnea, T and Turner, PJ and Harris, R and Mullins, GK and Bobowski, JS and Raudsepp, M and Liang, Ruixing and Bonn, DA and Hardy, WN},
  journal={Phys. Rev. B},
  volume={69},
  number={18},
  pages={184513},
  year={2004},
  publisher={APS}
}

@article{nishizaki_2003_vortex_JLTP,
  title={Vortex Phase Transition and Oxygen Vacancy in $\mathrm{YBa_2Cu_3O_y}$ Single Crystals},
  author={Nishizaki, Terukazu and Shibata, Kenji and Maki, Makoto and Kobayashi, Norio},
  journal={Jour. of Low Temp. Phys.},
  volume={131},
  number={5},
  pages={931--940},
  year={2003},
  publisher={Springer}
}

@article{gustavsson_2016_suppressing_Science,
  title={Suppressing relaxation in superconducting qubits by quasiparticle pumping},
  author={Gustavsson, Simon and Yan, Fei and Catelani, Gianluigi and Bylander, Jonas and Kamal, Archana and Birenbaum, Jeffrey and Hover, David and Rosenberg, Danna and Samach, Gabriel and Sears, Adam P and others},
  journal={Science},
  volume={354},
  number={6319},
  pages={1573--1577},
  year={2016},
  publisher={American Association for the Advancement of Science}
}

@article{bland_2025_millisecond_nature,
  title={Millisecond lifetimes and coherence times in 2D transmon qubits},
  author={Bland, Matthew P and Bahrami, Faranak and Martinez, Jeronimo GC and Prestegaard, Paal H and Smitham, Basil M and Joshi, Atharv and Hedrick, Elizabeth and Kumar, Shashwat and Yang, Ambrose and Pakpour-Tabrizi, Alexander C and others},
  journal={Nature},
  pages={1--6},
  year={2025},
  publisher={Nature Publishing Group UK London}
}

@article{munro_2002_weak_PRA,
  title={Weak-force detection with superposed coherent states},
  author={Munro, William J and Nemoto, Kae and Milburn, Gerard J and Braunstein, Sam L},
  journal={Phys. Rev. A},
  volume={66},
  number={2},
  pages={023819},
  year={2002},
  publisher={APS}
}

@article{belenchia_2018_quantum_PRD,
  title={Quantum superposition of massive objects and the quantization of gravity},
  author={Belenchia, Alessio and Wald, Robert M and Giacomini, Flaminia and Castro-Ruiz, Esteban and Brukner, {\v{C}}aslav and Aspelmeyer, Markus},
  journal={Phys. Rev. D},
  volume={98},
  number={12},
  pages={126009},
  year={2018},
  publisher={APS}
}

@article{bassi_2013_models_RMP,
  title={Models of wave-function collapse, underlying theories, and experimental tests},
  author={Bassi, Angelo and Lochan, Kinjalk and Satin, Seema and Singh, Tejinder P and Ulbricht, Hendrik},
  journal={Rev. of Mod. Phys.},
  volume={85},
  number={2},
  pages={471--527},
  year={2013},
  publisher={APS}
}

@article{wang_2014_measurement_NatureComm,
  title={Measurement and control of quasiparticle dynamics in a superconducting qubit},
  author={Wang, Chen and Gao, Yvonne Y and Pop, Ioan M and Vool, Uri and Axline, Chris and Brecht, Teresa and Heeres, Reinier W and Frunzio, Luigi and Devoret, Michel H and Catelani, Gianluigi and others},
  journal={Nat. Commun.},
  volume={5},
  number={1},
  pages={5836},
  year={2014},
  publisher={Nature Publishing Group UK London}
}

@article{nsanzineza_2014_trapping_PRL,
  title={Trapping a single vortex and reducing quasiparticles in a superconducting resonator},
  author={Nsanzineza, I and Plourde, BLT},
  journal={Phys. Rev. Lett.},
  volume={113},
  number={11},
  pages={117002},
  year={2014},
  publisher={APS}
}

@article{wells_2015_analysis_SciRep,
  title={Analysis of low-field isotropic vortex glass containing vortex groups in $\mathrm{YBa_2Cu_3O_{7-x}}$ thin films visualized by scanning SQUID microscopy},
  author={Wells, Frederick S and Pan, Alexey V and Wang, X Renshaw and Fedoseev, Sergey A and Hilgenkamp, Hans},
  journal={Sci. Rep.},
  volume={5},
  number={1},
  pages={8677},
  year={2015},
  publisher={Nature Publishing Group UK London}
}

@article{schlussel_2018_wide_PRApp,
  title={Wide-field imaging of superconductor vortices with electron spins in diamond},
  author={Schlussel, Yechezkel and Lenz, Till and Rohner, Dominik and Bar-Haim, Yaniv and Bougas, Lykourgos and Groswasser, David and Kieschnick, Michael and Rozenberg, Evgeny and Thiel, Lucas and Waxman, Amir and others},
  journal={Phys. Rev. Appl.},
  volume={10},
  number={3},
  pages={034032},
  year={2018},
  publisher={APS}
}

@article{straver_2008_controlled_APL,
  title={Controlled manipulation of individual vortices in a superconductor},
  author={Straver, Eric WJ and Hoffman, Jennifer E and Auslaender, Ophir M and Rugar, Daniel and Moler, Kathryn A},
  journal={Appl. Phys. Lett.},
  volume={93},
  number={17},
  year={2008},
  publisher={AIP Publishing}
}

@article{hoffman_2002_imaging_science,
  title={Imaging quasiparticle interference in $\mathrm{Bi_2Sr_2CaCu_2O_{8+\delta}}$},
  author={Hoffman, JE and McElroy, K and Lee, D-H and Lang, KM and Eisaki, H and Uchida, S and Davis, JC},
  journal={Science},
  volume={297},
  number={5584},
  pages={1148--1151},
  year={2002},
  publisher={American Association for the Advancement of Science}
}

@article{bahrami_2025_vortex_arXiv,
  title={Vortex motion induced losses in tantalum resonators},
  author={Bahrami, Faranak and Bland, Matthew P and Shumiya, Nana and Chang, Ray D and Hedrick, Elizabeth and McLellan, Russell A and Crowley, Kevin D and Dutta, Aveek and Bishop-Van Horn, Logan and Iguchi, Yusuke and others},
  journal={Phys. Rev. B},
  volume={113},
  number={5},
  pages={054505},
  year={2026},
  publisher={APS}
}

@article{eley_2016_decoupling_SST,
  title={Decoupling and tuning competing effects of different types of defects on flux creep in irradiated $\mathrm{YBa_2Cu_3O_{7- \delta}}$ coated conductors},
  author={Eley, S and Leroux, M and Rupich, MW and Miller, DJ and Sheng, H and Niraula, PM and Kayani, A and Welp, U and Kwok, WK and Civale, L},
  journal={Sup. Sci. and Tech.},
  volume={30},
  number={1},
  pages={015010},
  year={2016},
  publisher={IOP Publishing}
}

@article{olson_2004_vortices_PRL,
  title={Do vortices entangle?},
  author={Olson Reichhardt, CJ and Hastings, MB},
  journal={Phys. Rev. Lett.},
  volume={92},
  number={15},
  pages={157002},
  year={2004},
  publisher={APS}
}

@article{Dutta_2021_Evidence_PRB,
  title = {Evidence of zero-point fluctuation of vortices in a very weakly pinned $a$-MoGe thin film},
  author = {Dutta, Surajit and Roy, Indranil and Jesudasan, John and Sachdev, Subir and Raychaudhuri, Pratap},
  journal = {Phys. Rev. B},
  volume = {103},
  issue = {21},
  pages = {214512},
  numpages = {10},
  year = {2021},
  month = {Jun},
  publisher = {American Physical Society}
}

@article{wallraff_2003_quantum_nature,
  title={Quantum dynamics of a single vortex},
  author={Wallraff, Andreas and Lukashenko, A and Lisenfeld, J and Kemp, A and Fistul, MV and Koval, Y and Ustinov, AV},
  journal={Nature},
  volume={425},
  number={6954},
  pages={155--158},
  year={2003},
  publisher={Nature Publishing Group UK London}
}

@article{fruchter_1991_low_PRB,
  title={Low-temperature magnetic relaxation in $\mathrm{YBa_2Cu_3 O_{7-\delta}}$: Evidence for quantum tunneling of vortices},
  author={Fruchter, L and Malozemoff, AP and Campbell, IA and Sanchez, J and Konczykowski, M and Griessen, R and Holtzberg, F},
  journal={Phys. Rev. B},
  volume={43},
  number={10},
  pages={8709},
  year={1991},
  publisher={APS}
}

@article{hovhannisyan_2025_scanning_Comm_Materials,
  title={Scanning vortex microscopy reveals thickness-dependent pinning nano-network in superconducting niobium films},
  author={Hovhannisyan, Razmik A and Grebenchuk, Sergey Yu and Larionov, Semen A and Shishkin, Andrey G and Grebenko, Artem K and Kupchinskaya, Nadezhda E and Dobrovolskaya, Ekaterina A and Skryabina, Olga V and Aladyshkin, Alexey Yu and Dremov, Vyacheslav V and others},
  journal={Commun. Mat.},
  volume={6},
  number={1},
  pages={42},
  year={2025},
  publisher={Nature Publishing Group UK London}
}

@article{kuit_2008_vortex_PRB,
  title={Vortex trapping and expulsion in thin-film $\mathrm{Y Ba_2 Cu_3 O_{7-\delta}}$ strips},
  author={Kuit, Kl H and Kirtley, JR and Van Der Veur, W and Molenaar, CG and Roesthuis, FJG and Troeman, AGP and Clem, JR and Hilgenkamp, H and Rogalla, Horst and Flokstra, Jakob},
  journal={Phys. Rev. B},
  volume={77},
  number={13},
  pages={134504},
  year={2008},
  publisher={APS}
}

@article{blakemore_2021_search_PRD,
  title={Search for non-Newtonian interactions at micrometer scale with a levitated test mass},
  author={Blakemore, Charles P and Fieguth, Alexander and Kawasaki, Akio and Priel, Nadav and Martin, Denzal and Rider, Alexander D and Wang, Qidong and Gratta, Giorgio},
  journal={Phys. Rev. D},
  volume={104},
  number={6},
  pages={L061101},
  year={2021},
  publisher={APS}
}

@article{moore_2014_search_PRL,
  title = {Search for Millicharged Particles Using Optically Levitated Microspheres},
  author = {Moore, David C. and Rider, Alexander D. and Gratta, Giorgio},
  journal = {Phys. Rev. Lett.},
  volume = {113},
  issue = {25},
  pages = {251801},
  numpages = {5},
  year = {2014},
  month = {Dec},
}

@article{ranjit_2016_zeptonewton_PRA,
  title={Zeptonewton force sensing with nanospheres in an optical lattice},
  author={Ranjit, Gambhir and Cunningham, Mark and Casey, Kirsten and Geraci, Andrew A},
  journal={Phys. Rev. A},
  volume={93},
  number={5},
  pages={053801},
  year={2016},
  publisher={APS}
}

@article{ahn_2020_ultrasensitive_NatNano,
  title={Ultrasensitive torque detection with an optically levitated nanorotor},
  author={Ahn, Jonghoon and Xu, Zhujing and Bang, Jaehoon and Ju, Peng and Gao, Xingyu and Li, Tongcang},
  journal={Nat. Nano.},
  volume={15},
  number={2},
  pages={89--93},
  year={2020},
  publisher={Nature Publishing Group UK London}
}

@article{timberlake_2019_acceleration_APL,
  title={Acceleration sensing with magnetically levitated oscillators above a superconductor},
  author={Timberlake, Chris and Gasbarri, Giulio and Vinante, Andrea and Setter, Ashley and Ulbricht, Hendrik},
  journal={Appl. Phys. Lett.},
  volume={115},
  number={22},
  year={2019},
  publisher={AIP Publishing}
}

@article{wang_2024_mechanical_PRL,
  title={Mechanical detection of nuclear decays},
  author={Wang, Jiaxiang and Penny, TW and Recoaro, Juan and Siegel, Benjamin and Tseng, Yu-Han and Moore, David C},
  journal={Phys. Rev. Lett.},
  volume={133},
  number={2},
  pages={023602},
  year={2024},
  publisher={APS}
}

@article{afek_2022_coherent_PRL,
  title={Coherent scattering of low mass dark matter from optically trapped sensors},
  author={Afek, Gadi and Carney, Daniel and Moore, David C},
  journal={Phys. Rev. Lett.},
  volume={128},
  number={10},
  pages={101301},
  year={2022},
  publisher={APS}
}

@article{ahrens_2025_levitated_PRL,
  title={Levitated ferromagnetic magnetometer with energy resolution well below $\hbar$},
  author={Ahrens, Felix and Ji, Wei and Budker, Dmitry and Timberlake, Chris and Ulbricht, Hendrik and Vinante, Andrea},
  journal={Phys. Rev. Lett.},
  volume={134},
  number={11},
  pages={110801},
  year={2025},
  publisher={APS}
}

@article{werfel_2012_superconductor_SST,
  title={Superconductor bearings, flywheels and transportation},
  author={Werfel, FN and Floegel-Delor, Uta and Rothfeld, R and Riedel, T and Goebel, B and Wippich, D and Schirrmeister, P},
  journal={Sup. Sci. and Tech.},
  volume={25},
  number={1},
  pages={014007},
  year={2012}
}

@article{bernstein_2020_superconducting_SST,
  title={Superconducting magnetic levitation: principle, materials, physics and models},
  author={Bernstein, P and Noudem, J},
  journal={Sup. Sci. and Tech.},
  volume={33},
  number={3},
  pages={033001},
  year={2020},
  publisher={IOP Publishing}
}

@article{fung_2024_toward_PRL,
  title={Toward programmable quantum processors based on spin qubits with mechanically mediated interactions and transport},
  author={Fung, F and Rosenfeld, E and Schaefer, JD and Kabcenell, A and Gieseler, J and Zhou, TX and Madhavan, T and Aslam, N and Yacoby, A and Lukin, MD},
  journal={Phys. Rev. Lett.},
  volume={132},
  number={26},
  pages={263602},
  year={2024},
  publisher={APS}
}

@article{roda_2024_macroscopic_PRL,
  title={Macroscopic quantum superpositions via dynamics in a wide double-well potential},
  author={Roda-Llordes, Marc and Riera-Campeny, Andreu and Candoli, Davide and Grochowski, Piotr T and Romero-Isart, Oriol},
  journal={Phys. Rev. Lett.},
  volume={132},
  number={2},
  pages={023601},
  year={2024},
  publisher={APS}
}

@article{chang_1992_scanning_APL,
  title={Scanning Hall probe microscopy},
  author={Chang, AM and Hallen, HD and Harriott, L and Hess, HF and Kao, HL and Kwo, J and Miller, RE and Wolfe, R and Van der Ziel, J and Chang, TY},
  journal={Appl. Phys. Lett.},
  volume={61},
  number={16},
  pages={1974--1976},
  year={1992},
  publisher={American Institute of Physics}
}

@article{grosser_1997_detecting_arXiv,
  title={Detecting flux creep in superconducting YBCO thin films via damping of the oscillations of a levitating permanent magnet},
  author={Grosser, R and Martin, A and Niemetz, M and Pechen, EV and Schoepe, Wilfried},
  journal={arXiv preprint cond-mat/9712199},
  year={1997}
}

@article{Hardy_1993_Precision_PRL,
  title = {Precision measurements of the temperature dependence of \ensuremath{\lambda} in $\mathrm{YBa_2Cu_3 O_{6.95}}$: Strong evidence for nodes in the gap function},
  author = {Hardy, W. N. and Bonn, D. A. and Morgan, D. C. and Liang, Ruixing and Zhang, Kuan},
  journal = {Phys. Rev. Lett.},
  volume = {70},
  issue = {25},
  pages = {3999--4002},
  numpages = {0},
  year = {1993},
  month = {Jun},
  publisher = {American Physical Society}
}

@article{pesetski_2000_experimental_prb,
  title={Experimental study of the inductance of pinned vortices in superconducting $\mathrm{YBa_2Cu_3O_{7- \delta}}$ films},
  author={Pesetski, Aaron A and Lemberger, Thomas R},
  journal={Phys. Rev. B},
  volume={62},
  number={17},
  pages={11826},
  year={2000},
  publisher={APS}
}

@article{blatter_1994_vortices_RMP,
  title={Vortices in high-temperature superconductors},
  author={Blatter, Gianni and Feigel'man, Mikhail V and Geshkenbein, Vadim B and Larkin, Anatoly I and Vinokur, Valerii M},
  journal={Reviews of Modern Physics},
  volume={66},
  number={4},
  pages={1125},
  year={1994},
  publisher={APS}
}

@article{anderson_1964_hard_RMP,
  title={Hard superconductivity: theory of the motion of Abrikosov flux lines},
  author={Anderson, Philip W and Kim, YB},
  journal={Reviews of Modern Physics},
  volume={36},
  number={1},
  pages={39},
  year={1964},
  publisher={APS}
}

@article{grosser_1995_damping_APL,
  title={Damping of the oscillations of a permanent magnet levitating between high-Tc superconductors},
  author={Gro{\ss}er, Reiner and J{\"a}ger, Jan and Betz, J and Schoepe, Wilfried},
  journal={Appl. Phys. Lett.},
  volume={67},
  number={16},
  pages={2400--2402},
  year={1995},
  publisher={American Institute of Physics}
}

@article{grosser_2000_vortex_JLTP,
  title={Vortex Motion in Superconducting $YBa_2Cu_3O_{7-\delta}$ Inferred from the Damping of the Oscillations of a Levitating Magnetic Microsphere},
  author={Grosser, Reiner and H{\"o}cherl, Peter and Martin, Alfred and Niemetz, Michael and Schoepe, Wilfried},
  journal={Journal of Low Temperature Physics},
  volume={119},
  number={5},
  pages={723--742},
  year={2000},
  publisher={Springer}
}

@article{brown_2023_superfluid_PRL,
  title={Superfluid helium drops levitated in high vacuum},
  author={Brown, CD and Wang, Y and Namazi, M and Harris, GI and Uysal, MT and Harris, JGE},
  journal={Phys. Rev. Lett.},
  volume={130},
  number={21},
  pages={216001},
  year={2023},
  publisher={APS}
}

@article{timberlake_2024_linear_PRR,
  title={Linear cooling of a levitated micromagnetic cylinder by vibration},
  author={Timberlake, Chris and Simcox, Elliot and Ulbricht, Hendrik},
  journal={Physical Review Research},
  volume={6},
  number={3},
  pages={033345},
  year={2024},
  publisher={APS}
}

@article{druge_2014_damping_NJP,
  title={Damping and non-linearity of a levitating magnet in rotation above a superconductor},
  author={Druge, J{\'e}r{\'e}mie and Jean, C and Laurent, O and M{\'e}asson, MA and Favero, I},
  journal={New Journal of Physics},
  volume={16},
  number={7},
  pages={075011},
  year={2014},
  publisher={IOP Publishing}
}

\end{document}